\documentclass[a4paper,fleqn,usenatbib]{mnras}

\usepackage{newtxtext,newtxmath}

\usepackage[T1]{fontenc}
\usepackage{ae,aecompl}

\usepackage{graphicx}	
\usepackage{amsmath}	

\usepackage{amssymb}	
\usepackage{pdflscape}	
\usepackage{multirow}

\title[Red Be stars in the Magellanic Clouds?]{Red Be stars in the Magellanic Clouds?}

\author[Vieira K., Garc\'ia-Varela A., Sabogal B., R\'imulo, L. \& Hern\'andez, J.]{Katherine Vieira,$^{1,2}$\thanks{Contact email: katherine.vieira@uda.cl}
Alejandro Garc\'ia-Varela,$^{3}$
Beatriz Sabogal,$^{3}$
Leandro Rocha R\'imulo,$^{3}$
\newauthor and Jes\'us Hern\'andez$^{4}$
\\
$^{1}$Instituto de Astronom\'ia y Ciencias Planetarias INCT, Universidad de Atacama, Copayapu 485, Copiap\'o, Chile\\
$^{2}$Centro de Investigaciones de Astronom\'ia, Apartado Postal 264, M\'erida 5101-A, Venezuela\\
$^{3}$Universidad de los Andes, Departamento de F\'{\i}sica, Cra. 1 No. 18A-10, Bloque Ip, AA 4976, Bogot\'a, Colombia\\
$^{4}$ Instituto de Astronom\'ia, UNAM, Unidad Acad\'emica en Ensenada, Ensenada 22860, M\'exico}

\date{Accepted 2021 .... Received 2021 ...; in original form 2019 ...}

\pubyear{2021}

\begin{document}
\label{firstpage}
\pagerange{\pageref{firstpage}--\pageref{lastpage}}
\maketitle

\begin{abstract}
We revisit the subject of Be candidate stars towards the Magellanic Clouds, 
previously studied by the authors using SPM4 proper motions.
We obtain GAIA DR2 parallaxes and proper motions for 2357 and 994 LMC and SMC Be candidates, respectively.
Parallaxes and proper motions vs. color $V-I$ easily reveal the presence of the redder galactic contaminant foreground,
as concluded in our previous work, but this time we do find a few red Be candidates consistent with being true Magellanic objects. 
A membership assessment to each Magellanic Cloud is done for each Be candidate, 
based on the distribution of their parallaxes and proper motions.
From a compilation of published catalogues of spectroscopically confirmed emission line stars, 
we find that 41 (LMC) and 56 (SMC) of the Be candidates have shown the $H_{\alpha}$ line in emission.
Near-infrared IRSF $JHKs$ magnitudes are obtained for about 70\%  of the Be candidates with GAIA DR2 astrometric data.
Mid-infrared SAGE IRAC 3.6, 4.5, 5.8 and 8.0 \micron\,  magnitudes are obtained for about 85\% as well. 
After discarding possible Herbig Ae/Be stars, one LMC and three SMC B emission line confirmed stars show significantly redder optical, near- and mid-infrared colours than what has been typically measured for Classical Be stars, they are called red Be stars.
Taking into account that classical Be stars do not show these red colours, further studies about these four stars are needed in order to establish their
true nature and explain the observed red excess.

\end{abstract}




\section{Introduction}

Be stars are broadly defined as non-supergiant (luminosity class II to V) B-type stars that have 
or have had Balmer emission lines \citep{collins}.
The presence of a flattened circumstellar gaseous disk formed of material ejected from the star,
a dust-free Keplerian {\it decretion disc}, is currently the accepted explanation for some of the observed features in Be stars:
the UV stellar light is reprocessed in it and produces the emission lines, and the observed
IR excess and polarization result from the scattering of the stellar light by the disk (see \citet{review13} for details).
Several mechanisms have been proposed for the mass-ejection process that forms the disk, 
which are well constrained but not totally understood. In the so-called Classical Be (CBe)
it clearly comes from the rapid rotation of the star, 
probably along with other processes including non-radial pulsations and small-scale magnetic fields.
In binary stars the material is being accreted
by the companion of the Be star, generally a white dwarf. 

\citet{mennickent02} and \citet{sabogal05}, have identified a large number of Be candidate stars 
towards the Magellanic Clouds (MCs) based on their photometric variability in the OGLE-II $I$-band.
\citet{paul12} showed that this photometric method  is effective in the selection of Be candidates,
as their spectroscopic analysis found that most of the stars studied from a sample of such candidates
in both LMC and SMC, belong to early type stars with emission supporting circumstellar material.
However an enigmatic subgroup in the LMC sample was found and proposed as a possible subclass of stars that needed further analysis.
In \citet{vieira17}, we proved this subgroup was in fact Galactic foreground contamination,
as revealed by their SPM4 \citep{spm4} proper motions. In the SMC, only a few contaminants were found.

In this investigation, the results in \citet{vieira17}, are tested with GAIA DR2 \citep{gaia, gaiadr2},
that has significantly better proper motions and also measured parallaxes that were not available before.
The main result regarding the redder candidates being contaminants was indeed confirmed, 
but several red Be candidates emerged as true Magellanic stars. Though most of them are classified as Type-4
(see \citet{mennickent02} for this classification, according to their variability light curve morphology), 
and are considered by them and other references as
surely CBe stars, they do not occupy the infrared vs optical colours known so far for Be stars.

This paper go through the following topics: cleaning the Be candidates and matching with GAIA DR2 (sections 2 and 3),
statistical analysis of their parallaxes and proper motions and membership assessment (sections 4, 5 and 6),
crossmatch with spectroscopic catalogues to obtain the list of spectroscopically confirmed B emission line (BEL) stars (section 7), 
distribution of the light curve variability types of Be candidate and BEL stars (section 8),
near and mid-infrared photometry and finding redder than expected Magellanic confirmed BEL 
and Be candidate stars (section 9), some alternative explanations for the results (section 10), and  
conclusions (section 11).
Through this paper we use the following expressions: Be candidate stars, 
to indicate stars that were selected as such in the references consulted 
because of their photometric properties, in particular variability; 
and confirmed BEL, to indicate stars that have been previously reported as 
B stars with the $H_{\alpha}$ line in emission, and that are a subset of the Be candidate ones.

\section{Cleaning the B\lowercase{e} candidates} 

Be candidate stars for the LMC and SMC were obtained from \citet{sabogal05} and  \citet{mennickent02},
respectively, where a total of 2446 and 1019 candidates are listed.
In \citet{vieira17}, we had found a few repeated entries in each catalogue, with the same OGLE II identification, 
but in this reanalysis we found OGLE II has a slightly larger set of very similar entries, all correspond to pairs of stars,
located less than 2$\arcsec$ from each other, with very similar photometry ($<0.1$ magnitude differences) in the three $BVI$ bands.
These are repeated entries with bad enough astrometry to end up listed as different stars instead of the same
and in many cases they have different OGLE II identifications (which is based on their coordinates).
One entry per pair was kept (the first one listed), so the clean lists of Be candidates have 2393 and 1004 stars,
towards the LMC and SMC, respectively. 

\section{Crossmatching with GAIA DR2}\label{sec_xmatch}

\begin{figure}
    \includegraphics[width=\columnwidth]{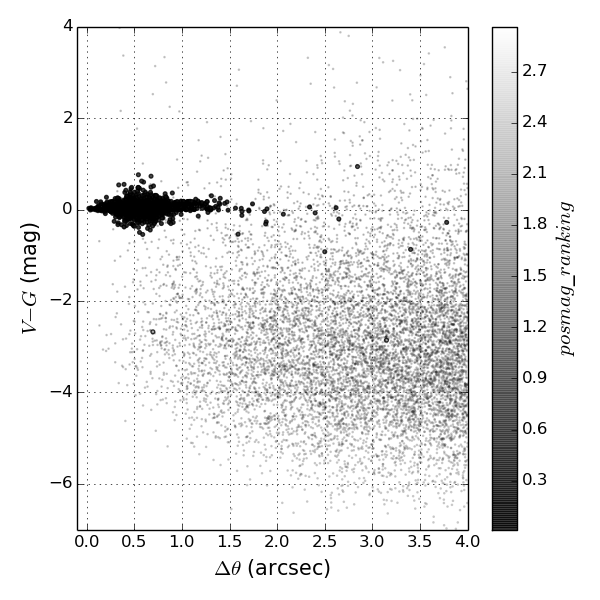}
    \caption{Matches between LMC Be candidates and GAIA DR2. This plot of $V-G$ vs. $\Delta\theta$, color-coded by {\it posmag\_ranking}, 
    for the GAIA DR2 entries chosen as matches for the Be candidates (filled circles), 
    suggests that except for a few outliers beyond $\Delta\theta\sim 2.7\arcsec$, the vast majority $(99.8\%)$ of these identifications are correct. 
    Positional-only matches within 4$\arcsec$ (light dots) are plotted as well, many of them are nearby fainter stars in GAIA DR2,
    and illustrate why {\it posmag\_ranking} is best to choose the true match in GAIA DR2.}
    \label{fig_histo_dtheta_dj}
\end{figure}

To properly crossmatch the Be candidates with GAIA DR2, we considered not only closeness in sky position
but also in photometry, since the GAIA DR2 $G$-band is close enough to the OGLE-II $V$-band,
as to serve to clean up possible misidentifications. GAIA DR2 is significantly deeper in magnitude,
therefore denser than OGLE-II, so there are chances for incorrect matches (see Figure \ref{fig_histo_dtheta_dj}). 
This approach is supported by the fact that among the only-positional matches up to 4$\arcsec$ 
a clear concentration of data points at $V-G\sim 0$ is visible.

In the SMC at least one GAIA DR2 match was found within 2$\arcsec$ that also looked close enough in photometry,
but in the LMC, for a few stars the best looking match was almost as far as 4$\arcsec$. 
The GAIA DR2 match was chosen as the star having the lowest value of
\begin{equation}
posmag\_ranking=\sqrt{\left(\frac{\Delta\theta}{\theta_{max}}\right)^2+\left(\frac{V-G}{1 \mbox{mag}}\right)^2}
\end{equation}
where 
$\Delta\theta$ is the sky angular separation in arcseconds between the (RA,DEC) coordinates listed for the Be candidates and the GAIA DR2 ones,
$\theta_{max}=2\arcsec$ for the SMC and  $\theta_{max}=4\arcsec$ for the LMC.
As the equation above suggests, $posmag\_ranking$ is defined to measure the closeness in both sky position and visual photometry.
All entries in the clean lists of Be candidates have a GAIA DR2 match, but we believe that those few with
$|V-G|\sim1$ mag or $\Delta\theta>3\arcsec$ are mismatches, as suggested by Figure \ref{fig_histo_dtheta_dj} . 
We do not discard any star for this reason, but we do keep in mind they could be mismatches.
We keep the $posmag\_ranking$ value for future reference.

\section{First analysis from parallaxes and proper motions}
\begin{figure}
    \includegraphics[width=\columnwidth]{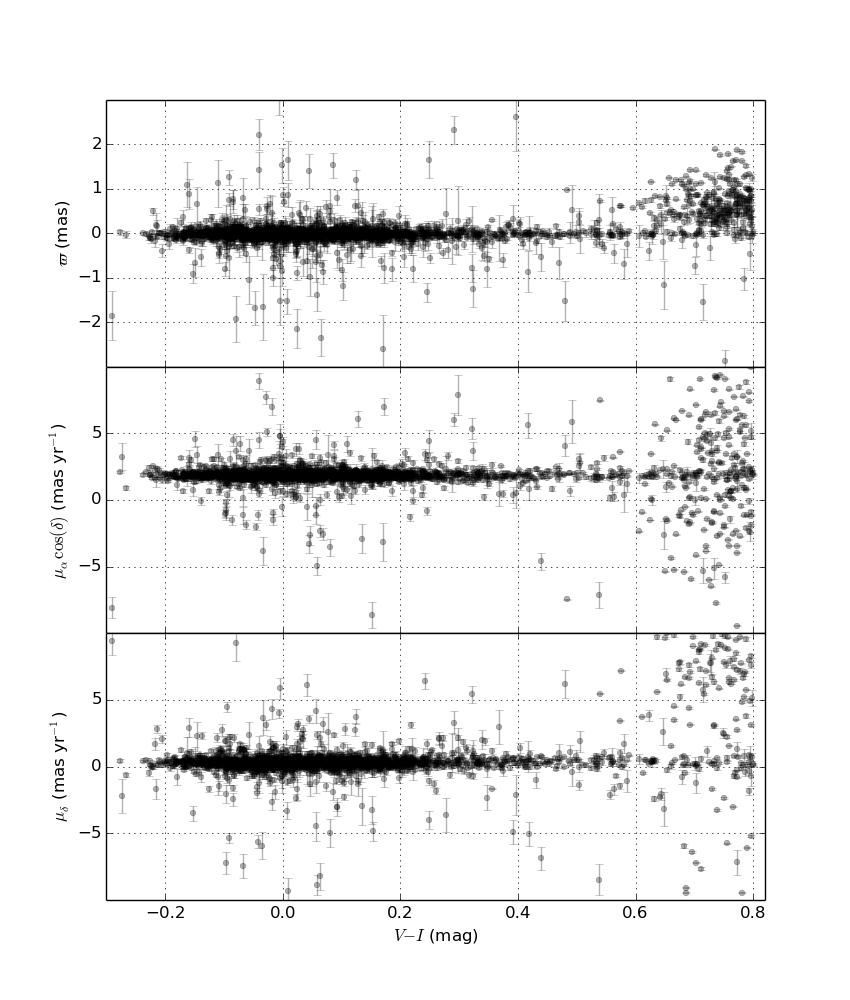}
    \includegraphics[width=\columnwidth]{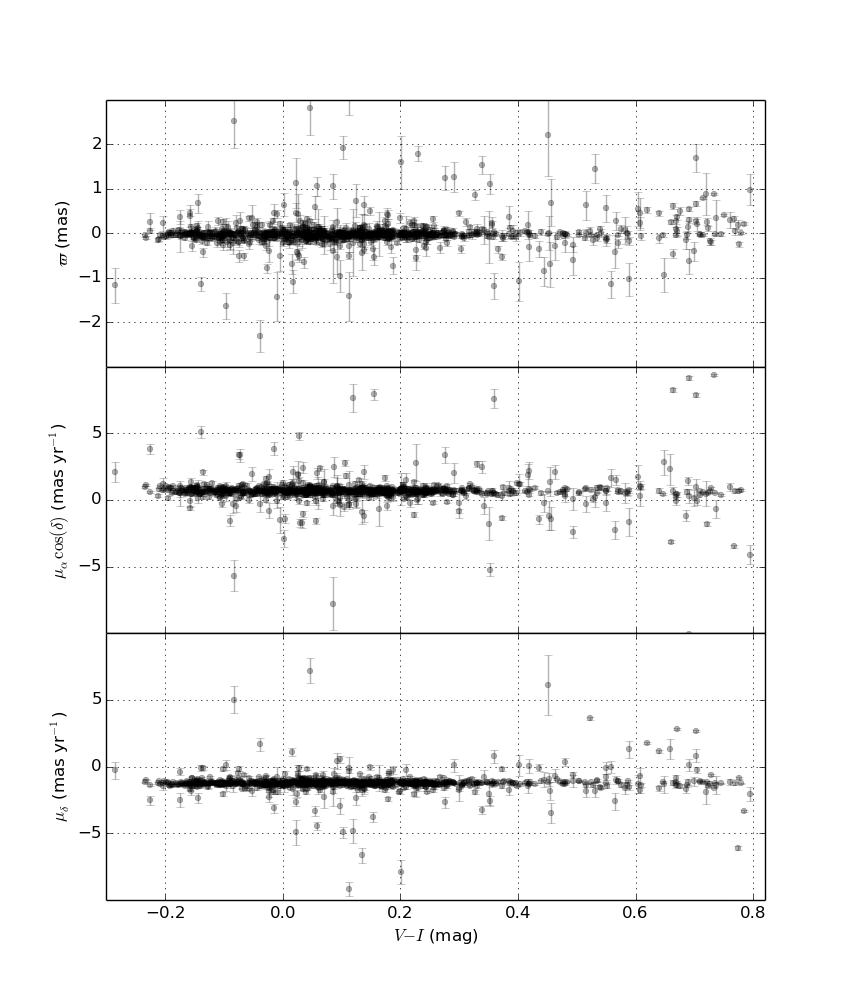}
    \caption{GAIA DR2 Parallaxes and proper motions with their respective error bars, 
    for the LMC (upper panel) and SMC (lower panel) Be candidates vs $V-I$ color.}
    \label{fig_data_vs_VI}
\end{figure}

Of the 2393 and 1004 GAIA DR2 matches for the LMC and SMC Be candidates, only 2357 and 994 respectively,
have parallaxes and proper motions measured in GAIA DR2, and with these datasets we will perform the
rest of the analysis. 
Figure \ref{fig_data_vs_VI}  clearly reveals the already known result from
\citet{vieira17}: there is a group of red stars ($V-I\gtrsim 0.6$) having parallaxes and proper motions that clearly puts them
as a foreground population residing in the Milky Way, while the bluer portion of
the candidates do exhibit the corresponding expected values for the Magellanic Clouds.
Nonetheless, not all of the red Be candidates belong to the foreground population,
there are $V-I\gtrsim 0.6$ objects that reside in the Magellanic Clouds.
This result arises clearly thanks to the manyfold improvement in proper motion precision and accuracy of GAIA DR2,
compared to SPM4's capabilities, that beats down errors to the point of allowing a clear separation of these two populations.
As expected, despite some overlap, stars belonging to the Magellanic Clouds do exhibit a
significantly larger concentration of their values of proper motion, which are further confirmed by parallaxes.
In fact, for a few blue Be candidates, it looks like their membership to the Magellanic Clouds could be reconsidered.
Not surprisingly, many of the candidates belonging to the Magellanic Clouds have larger errors, 
because of their significantly larger distance. But this feature is indeed a piece of valid information to consider
when assessing the membership of a Be candidate to these neighbour galaxies. 
As opposed to the usual approach where large error data are discarded in order to get clean samples,
in this work rejecting large-error data is in fact counterproductive.
On the other hand, since membership to a population is significantly decided by its closeness to the mean value of the population,
larger error data hinder finding that mean value with precision.

\section{Dealing with errors to find a good sample of Magellanic B\lowercase{e} candidates}
\begin{figure}
	\includegraphics[width=\columnwidth]{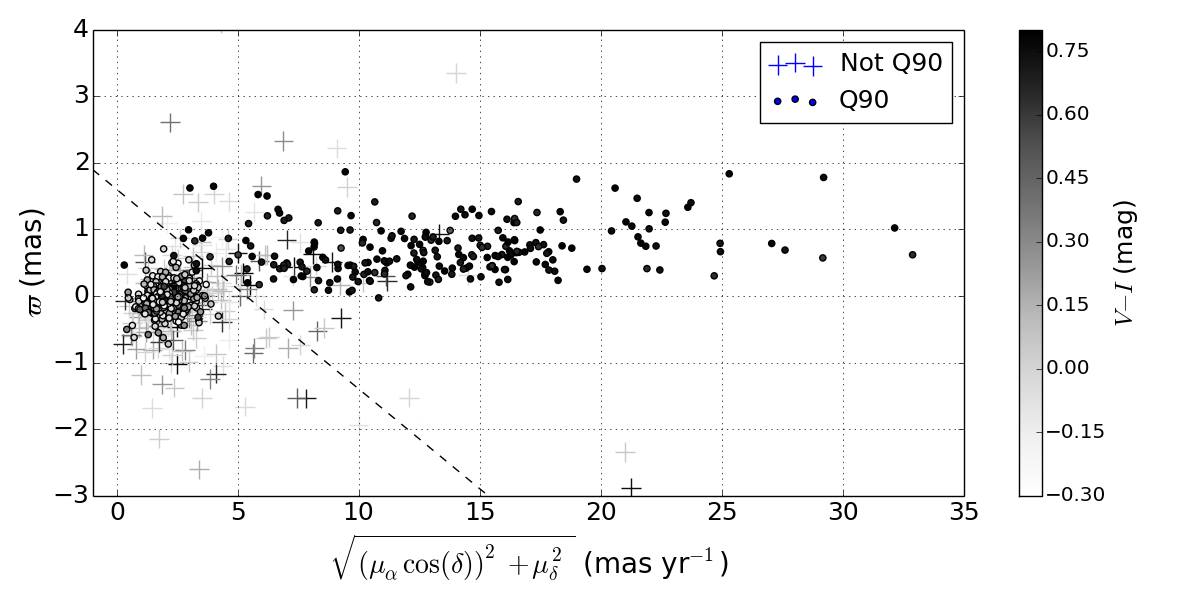}
    \caption{Plot corresponding to steps (ii) and (iii) for the LMC. Stars to the left of the dashed line are selected in step (iii) as sample LMC1.
    Stars to the right of the line are red in $V-I$ and have large total proper motions,
    they are mostly Milky Way foreground stars.  Data not in sample LMC Q90 (crosses) , 
    have large errors but they cluster around the LMC mean value, though with a much larger dispersion.}
    \label{fig_plx_vs_tpm_color_VI}
\end{figure}

The LMC and SMC absolute proper motions have been measured from the ground (e.g. \citealt{vieira10}) and
space (with GAIA DR2 itself, \citealt{gaiadr2MCpm}). Despite their small value as compared to Milky Way disk stars, for example,
measuring the angular displacement in the sky per unit time of the Magellanic Clouds  ($\sim$2 mas yr$^{-1}$) is a perfectly doable task for GAIA DR2,
given its proper motion mean precision of 0.06 to 0.15 mas yr$^{-1}$ for sources with $G<17$.
Parallax is a different issue, as the LMC and SMC distances ($\sim$ 50 and 60 kpc, respectively), translate into parallaxes
measurements of 20 and 16 $\mu$as, and GAIA DR2 parallax errors are about 40 to 90 $\mu$as for sources with $G<17$ \citep{luri19}.
 Given these circumstances, we decided to follow this approach, only to find the mean value of parallax and proper motions of the LMC and SMC:
\begin{enumerate}
\item Select the sample with all stars below the 90th quantile of proper motion error in both coordinates (from now on, samples LMC/SMC Q90).
\item Plot parallax $\varpi$ vs total proper motion $\sqrt{(\mu_\alpha\cos{\delta})^2+\mu_\delta^2}$, color coded by $V-I$ (see Figure \ref{fig_plx_vs_tpm_color_VI}).
\item From this plot, select the cluster of data corresponding to the LMC or SMC that is visibly separated from the foreground population
by their clustering in the plotted data and also the distribution of their $V-I$ colour (from now on, samples LMC1/SMC1).
For example, LMC1 is defined as all stars in LMC Q90 to the left of the dashed line plotted in Figure \ref{fig_plx_vs_tpm_color_VI}. Similarly happens for SMC1.
\item Apply a quantile-quantile or Q-Q plot on $\varpi, \mu_\alpha\cos{\delta}$ and $\mu_\delta$ for samples LMC1/SMC1,
to determine the mean and dispersion values for these data.
\item Adopt the obtained mean value for each measurement as the population mean of each Magellanic Cloud.
\end{enumerate}

\begin{figure}
	\includegraphics[width=\columnwidth,height=3in]{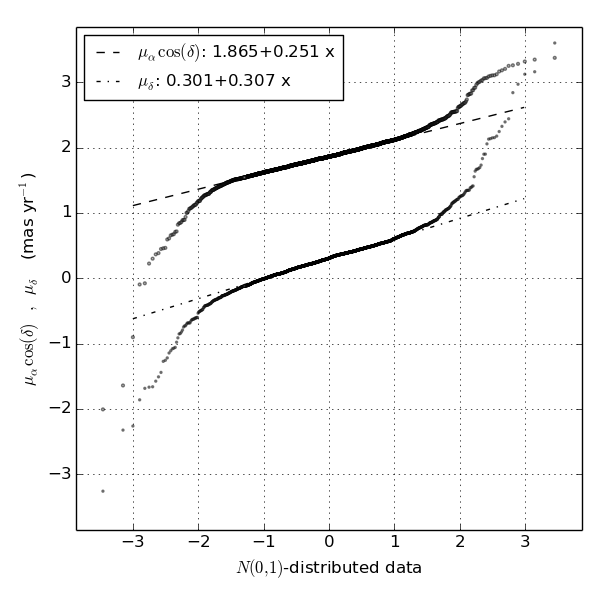}
    \caption{Q-Q plots results for the LMC proper motions, on the sample LMC1. 
    For the Q-Q fit, 10\% extreme data at each side (grey symbols) are not considered.}
    \label{fig_qq_lmc}
\end{figure}

Figure \ref{fig_plx_vs_tpm_color_VI} illustrates steps (ii) and (iii) for the LMC, and the separation of the two populations is clear and evident,
when the color information is considered.
In step (iv), the Q-Q plots are computed with respect to a standard normal distribution $N(0,1)$, 
so if the data do follow a normal distribution, the Q-Q plot will look like a straight line where the
y-intercept indicates the mean value and the slope corresponds to the dispersion. As visible in 
Figure \ref{fig_qq_lmc}, showing the Q-Q plots of $\mu_\alpha\cos{\delta}$ and $\mu_\delta$ for the LMC1 sample,
the 10\% extreme data points at each side were not considered for the fit. 
It is visible that those extreme points deviate visibly from the straight line (the tails of the distribution are heavier than normal),
which is not entirely unexpected. Besides some contamination from foreground stars, the observed dispersion is mostly caused by
error measurements and the samples studied have data of various quality. 
Since the points used in the fit do follow very closely a linear relation and the LMC1 and SMC1 datasets have
symmetrical distributions in the Q-Q plots, the obtained mean values are very robust. 
Similarly happens for parallaxes in both LMC and SMC.
Resulting mean and dispersion values are listed in Table \ref{tab_qq_results}.

\begin{table}
	\centering
	\caption{Mean and dispersion values of parallax and proper motions for LMC and SMC  Be candidates, as obtained from Q-Q plots
	on samples LMC1 and SMC1.}
	\label{tab_qq_results}
	\begin{tabular}{crccc} 
		\hline
		 & Sample & $\varpi$ & $\mu_\alpha\cos{\delta}$ & $\mu_\delta$  \\
		& size & (mas) & (mas yr$^{-1}$) & (mas yr$^{-1}$) \\ 
		\hline
		LMC1 & 1856  &-0.015 $\pm$ 0.081 & 1.865 $\pm$ 0.251 & \;0.301 $\pm$ 0.307    \\
		SMC1 & 872  & -0.023 $\pm$ 0.075 & 0.680 $\pm$ 0.202 & -1.220 $\pm$ 0.141 \\
		\hline
	\end{tabular}
\end{table}

It can be noticed that the mean parallax value is negative for both Clouds, and significantly enough compared to their dispersions.
GAIA DR2 parallaxes are known to have an offset towards smaller than real values (GAIA DR2 $-$ true value $<0$), for an amount 
that has been computed, among others, to be $-0.082\pm0.033$ using eclipsing binaries \citep{stassun18},
$-0.0528\pm 2.4$ (stat.)$\pm 1$(syst.) mas using red giant branch and Helium-burning red clump stars \citep{zinn19}, 
and the official value using quasars given by the GAIA Collaboration of -0.030 mas \citep{luri}.
In the LMC such offset has been found to have in fact systematic periodic variations (see Figure 13 in \citealt{arenou18}). 
Our results are within this range of biases, as compared to the expected parallaxes of 0.020  and 0.016 mas for the LMC and SMC,
respectively. It is not of concern what the mean values are for each Cloud, as we only use them to separate Magellanic Be candidates
form foreground population, based on the clustering of the data. 

As for the measured dispersion, in all the three kinds of data, within each Cloud, the intrinsic expected cosmic dispersion 
is substantially smaller than the observed one. For the young population in the Magellanic Clouds, velocity dispersion has been measured 
to be as low as 6 km s$^{-1}$ \citep{gyuk00}. A 15 km s$^{-1}$ velocity dispersion at 50(60)  kpc  
yields a proper motion dispersion of 0.06(0.05) mas yr$^{-1}$. 
Therefore the observed variance is caused dominantly by measurement errors.

\section{Data standardization and membership assessment to the LMC/SMC}
\begin{figure*}
	\includegraphics[width=7in]{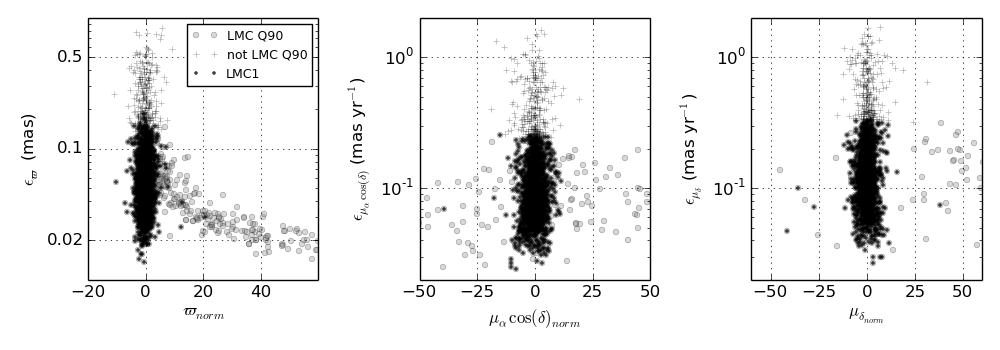}
    \caption{Individual  data errors vs normalized data 
    for LMC Be candidates. Despite their larger errors,
    data not in sample LMC Q90 (crosses)  
    cluster around the LMC mean values (zero) in normalized variables, the same way best measured LMC1 data do. 
    Discarding larger error data from start would systematically bias the obtained samples of Magellanic Clouds stars.}
    \label{fig_inderr_vs_normdata}
\end{figure*}

When the dispersion of a univariate sample of data is dominated by the measurement  errors,  
the non-constant scale of these data produce problems trying to analyze them. 
In order to obtain a unique scale for the data, the standardization of data \citep[p. 193]{Fei} is widely used
in the statistics and astrostatistic context, using the formula given by
\[
\mbox{\it datum}_{\mbox{\;\it norm}} = \frac{\mbox{\it datum}\; - \mbox{\it population mean}}
{\mbox{\it datum}_{\mbox{\;\it error}}}.
\]
This standardization process should yield a normal standard $N(0,1)$ distribution.   
Extended explanations of this method can be found in \citet[p. 19]{booknorm}, \citet[p. 50]{Hei} and \citet[p. 19]{Heu}.

If the dispersion $S$ of the normalized data is significantly $\neq$ 1,
a scaling of the errors {\it may} solve the problem, revealing that original quoted errors were under- ($S>1$) or overestimated ($S<1$).
This works and must work for all individual errors, whether large or small. 
We computed normalized data for all the LMC and SMC stars,  using the mean values obtained from LMC1 and SMC1, respectively.
A plot of the individual data error vs the normalized data,
as seen in Figure \ref{fig_inderr_vs_normdata} for the LMC stars, reveals in first place
the expected concentration around zero for stars that truly belong to the LMC, 
while foreground Galactic stars behave visibly different, both in mean and dispersion of the normalized data,
foreground stars have a huge dispersion that is in fact mostly cosmic. 
Sample LMC Q90 containing stars well measured in proper motions, include both Magellanic Clouds stars and also
Milky Way ones, the latter easily visible with large values of normalized data. Sample LMC1 concentrates around zero,
and so does Sample "not LMC Q90", which highlights once again how using normalized data allows us to keep these data that
behaves like members of the Magellanic Clouds, despite having larger individual errors. 
These larger error data follow the mean of the LMC though with a larger dispersion than
for the LMC1 sample, but still significantly smaller than the dispersion of foreground stars.  Similarly happens with the SMC data
though the foreground population is significantly lesser, as we already know.
Q-Q plots on the normalized data for the LMC1 and SMC1 samples, that we consider are constituted by single LMC and SMC populations respectively,
show that for these samples the individual errors are all underestimated by at least the corresponding dispersions found (results in Table \ref{tab_qq_norm_results}). 
As before, the 10\% extreme data showed a higher slope indicative of a higher scaling needed to {\it standardize} the data.
The true underestimation factor may not be as uniform as a single scale value
even across the LMC1 and SMC1 samples, and it is certainly not enough for the data outside them,
that could suffer from even larger understimations\footnote{With the largest possible LMC/SMC population sample,
the slopes of the Q-Q plots at each normalized datum can in fact be taken as the scaling factor for the corresponding individual error.
That would effectively fully {\it standardize} that sample.}.
This - again - is not entirely unexpected, as the proper GAIA consortium has warned of error underestimation in the GAIA DR2 data \citep{gaiadr2}.

\begin{table}
	\centering
	\caption{Mean and dispersion values of normalized parallax and proper motions (which are unitless), as obtained from Q-Q plots,
	for samples LMC1 and SMC1.}
	\label{tab_qq_norm_results}
	\begin{tabular}{crccc} 
		\hline
		 & Sample & $\varpi_{\mbox{\it norm}}$ & $\mu_\alpha\cos{\delta}_{\mbox{\it norm}}$ & $\mu_{\delta_{\mbox{\it norm}}}$ \\  
		 & size &  &  &  \\
		\hline
		LMC1 & 1856 & -0.036 $\pm$ 1.522 & -0.041 $\pm$ 2.865 & 0.038 $\pm$ 2.707  \\
		SMC1 & 872 & -0.042 $\pm$ 1.534 & \;0.039 $\pm$ 1.989 & 0.020 $\pm$ 1.801 \\
		\hline
	\end{tabular}
\end{table}

\subsection{Membership assessment}
As the scaling factor for the data individual errors is not uniform across the Magellanic data, using the obtained scales with the LMC1 and SMC1 samples,
to {\it standardize} the whole normalized datasets, and use a normal standard probability distribution $N(0,1)$ as a model of the Magellanic populations on such data,
is not entirely correct. And though these standard data help not to loose real Magellanic stars because of their individual larger errors/dispersion,
there is also the issue of how to model the distribution of the foreground Galactic population in those same variables.
The general approach is to model the expected frequency of stars for each of the populations present in the chosen variables, 
and then the probability to belong to a given population is the ratio of its expected frequency over the sum of the same for all the populations present.
A model for the latter is not an easy task because of varying systematic trends in proper motion data caused by e.g. solar motion or galactic rotation,
also clusters populations may distribute themselves following a non-normal or non-symmetric distributions.
Parametric (e.g. the seminal paper by \citealt{vasilevskis}) and non-parametric approaches (e.g. \citealt{galadi}) have been devised to treat this issue.

\begin{figure}
	\includegraphics[width=\columnwidth]{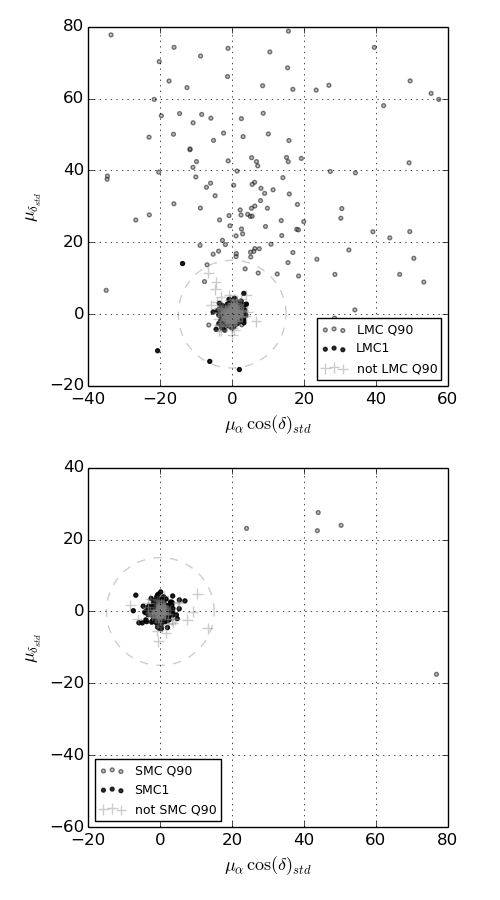}
    \caption{Standardized proper motions of the Be candidates towards the LMC (upper panel) and SMC (lower panel).
    In the SMC, the cut at radius$=$15 from the origin (sample SMC2) cleanly separates Magellanic from foreground population.
    This same limit applied to the LMC (sample LMC2) does not work as neatly due to larger and overlapping foreground contamination.}
    \label{fig_norm_pm}
\end{figure}

\begin{figure}
	\includegraphics[width=\columnwidth]{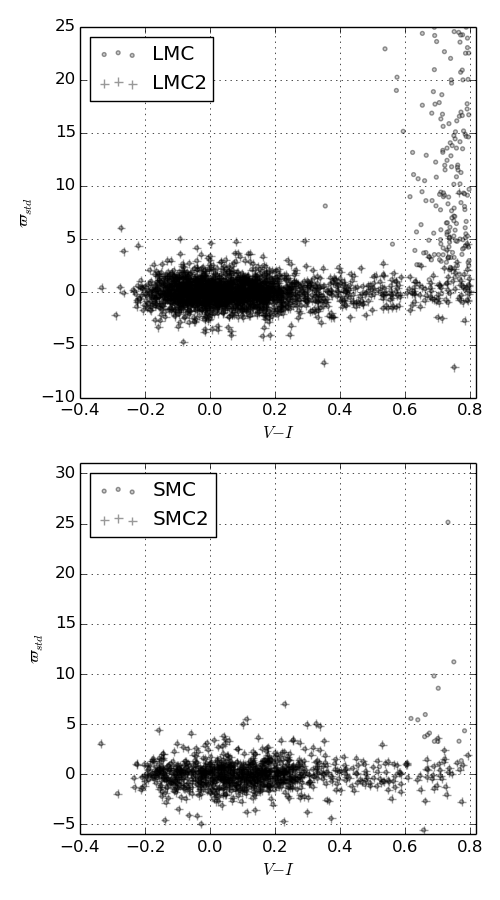}
    \caption{Standardized parallaxes of the Be candidates towards the LMC (upper panel) and SMC (lower panel) vs $V-I$.
    Black crosses correspond to samples LMC2 and SMC2, chosen solely from their standardized proper motions.}
    \label{fig_norm_plx}
\end{figure}

In the SMC, where we already know that contamination is less, a plot of the standardized proper motions for all the SMC data,  
shows a clear concentration of data points within a unitless  distance of 15 from (0,0), 
that we call sample SMC2 (976 stars), well isolated from a few scattered stars
at distances larger than 20 from the origin. This same cut does not look as clean in the LMC (sample LMC2, 2116 stars) where contamination is larger and Milky Way stars
overlap more with the LMC population (see Figure \ref{fig_norm_pm}). Interestingly, the above cut in standardized proper motions
is enough to automatically select stars with standardized parallaxes close to zero, as seen in Figure \ref{fig_norm_plx}.
The other way around, that is selecting first by standardized parallaxes is more visibly affected by foreground contamination.

\begin{figure}
	\includegraphics[width=\columnwidth]{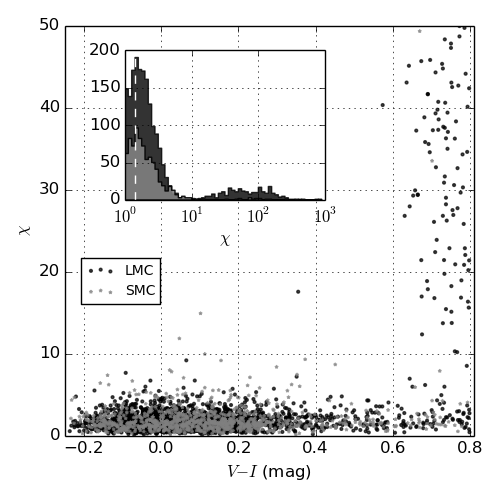}
    \caption{Plot of $\chi$ values vs $V-I$ for the LMC (black) and SMC (grey) Be candidates.
    From the distribution of the data points for both Clouds, we put $\chi=10$ as limit to
    separate Magellanic populations from Galactic ones. Inset figure shows the histogram in 
    a log-scale of $\chi$ values, that encompasses all Be candidates,
   Galactic stars can be seen occupying a very extended range in $\chi$. 
   The vertical white dashed lines marks $\chi=\sqrt{2}$, the mode for a $\chi$ distribution with 3 degrees of freedom.}
    \label{fig_std_histo}
\end{figure}

In any case, Magellanic stars cluster around zero simultaneously in standardized parallax and proper motions.
This leads us to define a new variable $\chi$, as follows:
\[
\chi = \sqrt{\left(\varpi_{\mbox{\it std}}\right)^2+\left(\mu_\alpha\cos(\delta)_{\mbox{\;\it std}}\right)^2+\left({\mu_\delta}_{\mbox{\;\it std}}\right)^2}
\]
which, should follow closely a $\chi$ (Chi) distribution with 3 degrees of freedom (\citet{statsbook}). 
Such probability distribution has a mode at $\chi=\sqrt{2}$,
and $\chi\leq 3.389$ should enclose enclose 99\% of the Magellanic population, if all standardized data truly followed $N(0,1)$ distributions
and were also free of contamination .   
Figure \ref{fig_std_histo} plots $\chi$ vs $V-I$ for the LMC and SMC data,  and from the distribution
of both the bluer Magellanic population and the redder Galactic one, we decide to draw a limit between the two at $\chi=10$.
We believe that incomplete standardization of the most extreme data within the Magellanic populations has caused them to stray beyond
the expected values for the $\chi$ distribution. Contamination can also play a role, but we expect it to be minimal at low values of $\chi$,
as shown by the histogram of $\chi$ for the whole LMC and SMC data (see inset in Figure \ref{fig_std_histo}) 
that clearly shows Magellanic stars clustering around  $\chi=\sqrt{2}$, and a secondary peak much farther away obviously caused by Milky Way stars. 
In any case, our standardized data are good in a relative sense, since Galactic foreground stars distribute themselves visibly far away from
the cluster of Magellanic data points around zero. In other words, the data themselves have indicated us where to do
the final cut to select the best possible single population LMC and SMC samples,  that is all stars with $\chi<10$, 
which yields samples LMC3 (2109 stars) and SMC3 (974 stars).

\begin{figure}
	\includegraphics[width=\columnwidth]{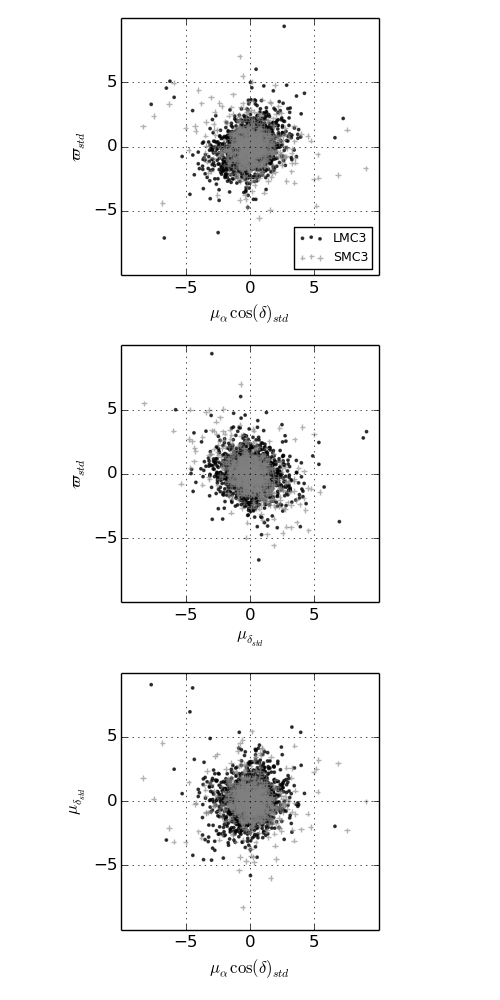}
    \caption{Correlations in standardized parallaxes and proper motions of the Be candidates in the LMC3 (black) and SMC3 (grey) samples. 
    Parallax and proper motions look a bit correlated though proper motions not among themselves.}
    \label{fig_std_correlations}
\end{figure}

Using $\chi$ as we just did, implies we are considering the three types of standardized data as independent.
We must mention at this point, that we did observe some slight correlation between the
normalized parallax and proper motions, though not among the proper motions.
These correlations were visibly reduced - but still a bit visible - in the standardized data 
(because the scaling did not work for all data, as we explained above, see Figure \ref{fig_std_correlations}).
We believe our assessment is nonetheless precise enough for the purpose of our investigation,
correctly separate Magellanic from Galactic stars.

When plotting the original $\varpi, \mu_\alpha\cos(\delta)$ and $\mu_\delta$ data vs $V-I$ color for the LMC3 and SMC3 samples, there appear to be some
blue outliers that do not concentrate around the LMC/SMC mean value as much as the rest of the data. 
We have concluded that their error bars are substantially larger than what is quoted by GAIA DR2, and that would explain
their apparent random distant location from the mean. We do in fact found that blue stars in the Be candidates have systematically larger errors than
red ones in GAIA DR2, but it could be caused precisely by the dominance of LMC/SMC stars in that color, which because of their distance
are more difficult  to measure.

\section{\textbf{Crossmatch with spectroscopic catalogues}}\label{sec_confbe}
Our lists of Be candidates were crossmatched with several previous publications 
and online databases of spectroscopically confirmed Be stars towards the Magellanic Clouds area, listed in Table \ref{tab_crossbe}. 
Despite these stars are called Be stars in those works, they are in many cases B emission line (BEL) stars at the time of observations, 
and not necessarily CBe. The crossmatch radius used was 2.5$\arcsec$, but the vast majority of our Be candidates were matched within 1$\arcsec$.
A few of our Be candidates appear simultaneously in several of these catalogues, because they have common stars
among them. Only one Be candidate in the SMC (ID=3), was matched to two different confirmed BEL stars,
{\it smc1-10} and {\it smc2-12} in \citet{paul12}, which in that publication
have the same coordinates and OGLE data, but different radial velocity measurements, though similar within their error bars.
The stars also have very similar spectral range. Whichever is the correct match, both are confirmed BEL stars.

\begin{table*}
	\centering
	\caption{Catalogues of spectroscopically confirmed BEL stars crossmatched with our Be candidates in the Magellanic Clouds.}
	\label{tab_crossbe}
	\begin{tabular}{cccc} 
		\hline
		Reference &  Galaxy & Tables and Comments about BEL confirmed stars & Number of stars matched\\
		\hline
		\citet{martayan07} & SMC & Table 3 (N\_Star=1 in electronic table) & 7\\
		\citet{martayan10} &  SMC & Tables C1, C2 (Code=1 in electronic table) & 10\\ 
		\citet{paul12}  & LMC, SMC & Tables 2 to 7: SpT=B0 to B9 & 2, 31 \\
		\citet{reid12}  & LMC & Tables A1, A2: SpType=Be II to V and H$_{\alpha}$ in Comments & 35 \\
		\citet{cieslinski13}  & LMC, SMC & Table 1:  All except  Notes=2,3,6,10,11,14 and 15 & 4, 6  \\
		\citet{sheets13} & SMC & Table 4: those with SpT=OBe  & 6 \\
		Be Stars Spectra database & SMC & Radial velocity $>$ 100 km s$^{-1}$, located in the MCs area & 10 \\ 
		(\url{http://basebe.obspm.fr})  &&& \\ 
		\hline
	\end{tabular}
\end{table*}

A total of 41 of our LMC Be candidates have a confirmed BEL star match,
and so do 56 of our SMC Be candidates. All of the confirmed BEL stars belong to the Magellanic Clouds according to
our criteria $\chi<10$ (LMC3/SMC3 samples), except one (ID=462 in LMC) that has no GAIA DR2 data and therefore no value to decide its
membership to the LMC. Also, all of the confirmed BEL stars have $|V-G|<1$ and $\Delta\theta<2.7\arcsec$, therefore
their GAIA DR2 data are most surely true matches. 

Examining the confirmed BEL stars sample in more detail, we find they all have  $\chi<4.5$ except
two stars in the SMC: ID = 55 ($\chi$=7.4) and  ID = 798 ($\chi$=9.2), which are both red in optical $V-I>0.35$, and both have the largest errors and deviations from the mean value in all parallaxes and proper motions in this sample. Star ID = 55 has a duplicate flag in GAIA DR2.
Among the SMC confirmed BEL stars, the following IDs have a duplicate flag in GAIA DR2: 55, 85, 88, 89, 103, 284 and 331,
and the only one LMC confirmed BEL star with such a flag is ID = 684.

A crossmatch with SIMBAD was performed as well, to explore how our Be candidates were classified in it.
SIMBAD is a highly non-uniform database, that contains a lot of information for some stars but none about many others.
A cross-match within 3$\arcsec$ with our lists of Be candidates showed in several cases multiple matches.
Keeping the closest match, we found that most of the stars were classified (under Object Type) as variable, 
a portion of them as Be stars, others as Long Period, Ellipsoidal and Eclipsing variables, etc.
Verifying the certainty of these classifications is beyond the scope of this paper,
it simply illustrates that a portion of our Be candidates can in fact be other types of variables,
which is not entirely unexpected.

\section{OGLE-based variability classification of the B\lowercase{e} candidate and confirmed BEL stars}
Our Be candidates have a classification in Types 1 to 4, proposed by \citet{mennickent02},
based on their OGLE $I$-band light curves morphology: Type-1 stars show outbursts,
Type-2 stars show high and low states, Type-3 stars show periodic variations and
Type-4 show stochastic variations. There are stars that behave like Type-1 and
Type-2 simultaneously and are classified as Type 1/2. In our study only the LMC has the latter.

\begin{table}
	\centering
	\caption{Variability types of confirmed BEL and Be candidate stars in the LMC3 and SMC3 samples.
	Percentages with respect the total sample being considered is shown below each line.}
	\label{tab_types}
	\begin{tabular}{crrrrrr} 
		\hline
		\multicolumn{7}{c}{confirmed BEL stars} \\ \hline
		Sample/Type & 1 & 1/2 & 2 & 3 & 4 & Total\\  		\hline
		LMC3 & 8 & 4  & 3 & 7 & 19 & 41  \\
		   & 20 \% & 10 \% & 7 \% & 17 \% & 46 \% & 100 \% \\
		SMC3 & 16  & 0  & 13  & 6  & 21 & 56 \\
		          & 29 \% & 0 \% & 23 \% & 11 \% & 37 \% & 100 \%\\		\hline
		\multicolumn{7}{c}{Be candidate stars} \\ \hline
		Sample/Type & 1 & 1/2 & 2 & 3 & 4 & Total\\  		\hline
		LMC3 & 554  & 95  & 148  & 142 & 1170  & 2109  \\
		            & 26 \%  & 5 \% & 7 \% & 7 \% & 55 \% & 100 \% \\
		SMC3 & 102  & 0 & 149  & 77  & 646 & 974 \\	
		            & 11 \% & 0 \% & 15 \% & 8 \% & 66 \% & 100 \%	\\ \hline
	\end{tabular}
\end{table}

Table \ref{tab_types} shows how many stars per Type are in the LMC3/SMC3 samples and also how many of these are confirmed BEL stars.
It reveals that confirmed BEL  stars can be found in all types, but they are more common in Type-4.
In the SMC, Type-1 confirmed BEL stars are almost as many as Type-4 ones, while in the LMC
Type-1 ones amount to less than half of the Type-4 ones. 
In \citet{sabogal05}, it is suggested that the mechanism behind the variability of Type-1 stars
could depend on metallicity: in low metallicity stars, rotation probably combined with
non-radial pulsations may be the outbursts main driver, while stellar winds would have
a reduced contribution. Our results with the SMC vs LMC confirmed BEL stars are in line with this idea.

When considering the full LMC3/SMC3 samples (that include the confirmed BEL stars), 
such proportions change only in the SMC, where Type-1 stars reduce even more their proportion to Type-4 ones.
It must be taken into account that the Be candidates samples may have contamination
from other non-Be variable stars, therefore, it is not unexpected to find some differences.

\section{Red Magellanic B\lowercase{e} stars?}\label{sec_red_be}

Like in our previous investigation \citep{vieira17}, we cross-matched our clean lists of LMC 2393 and 1004 
SMC Be candidates with a GAIA DR2 counterpart, with the IRSF catalogue by \citet{kato07}. An easier tracking of all names and sizes of the samples used in this work can be seen in Table \ref{tab_all.samples}.
GAIA DR2 2000.0 epoch coordinates were matched to the coordinates published by IRSF, and we found that true matches have
angular distances distributed below 0.4$\arcsec$, a rather small distance, 
within which 1521 (64\% of the LMC sample) and 737 (73\% of the SMC sample) stars were matched 
between these catalogues\footnote{These samples are a bit larger than the ones we found when crossmatching the SPM4 catalogue with the Be candidates
in \citet{vieira17}, then we found 1188 and 619 matched stars in the LMC and SMC, respectively.}.
We did the same  with the SAGE IRAC catalogues for the LMC
and SMC by \citet{meixner06} and \citet{gordon11}, respectively. 
The LMC IRAC has 6,398,991 entries and the SMC one has 2,015,403 entries.
These two catalogues were checked for repeated IDs or close entries within 2$\arcsec$ on the sky, and none were found. 
True matches between our stars and these catalogues were all within 1$\arcsec$, 
and the obtained matches, all singles, were 1882 in the LMC (79\% of the sample) and 869 in the SMC (87\% of the sample).
Not unexpectedly, some stars are missing either near-infrared or mid-infrared photometry or both.

\begin{figure*}
	\includegraphics[width=6in,height=4.5in]{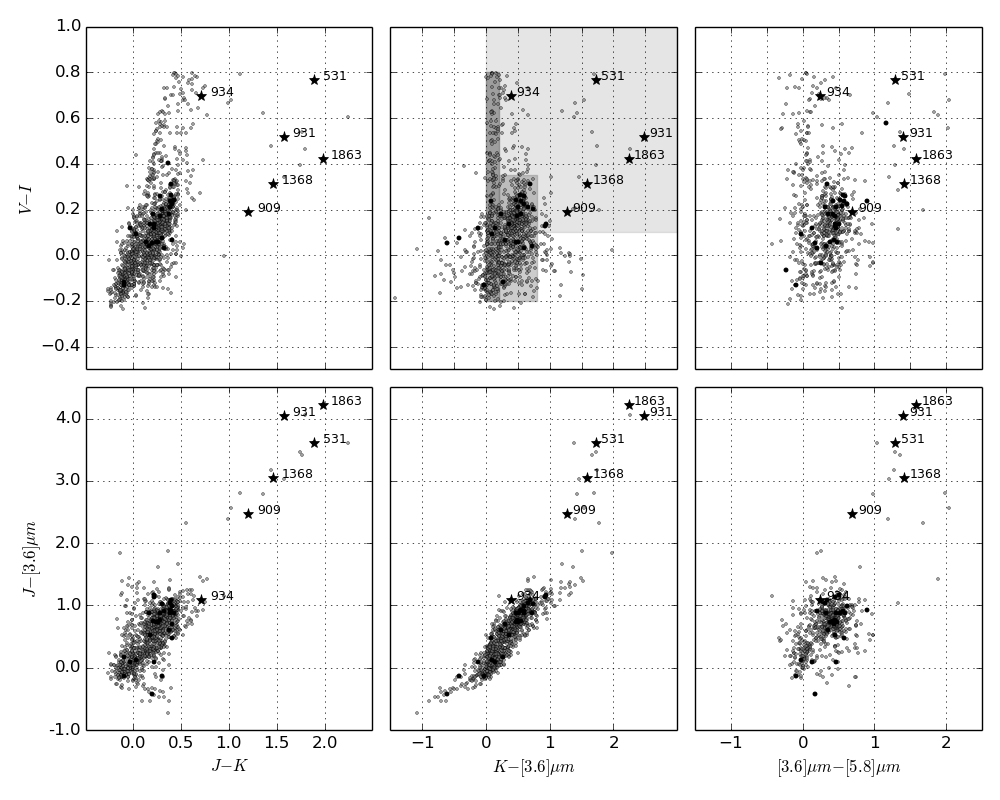}
    \caption{Optical, near- and mid-infrared colour-colour diagrams for the LMC3 sample. BEL confirmed stars are highlighted in darker symbols.
    Black dark star symbols are what we call red Be stars.
    Some stars (including BEL confirmed ones) in the LMC3 sample are missing either IRSF or IRAC photometry. The upper middle panel shows in decreasing shade gray colour, the regions corresponding to the subsets A, B and C, respectively.}
    \label{fig_cc_lmc}
\end{figure*}
\begin{figure*}
	\includegraphics[width=6in,height=4.5in]{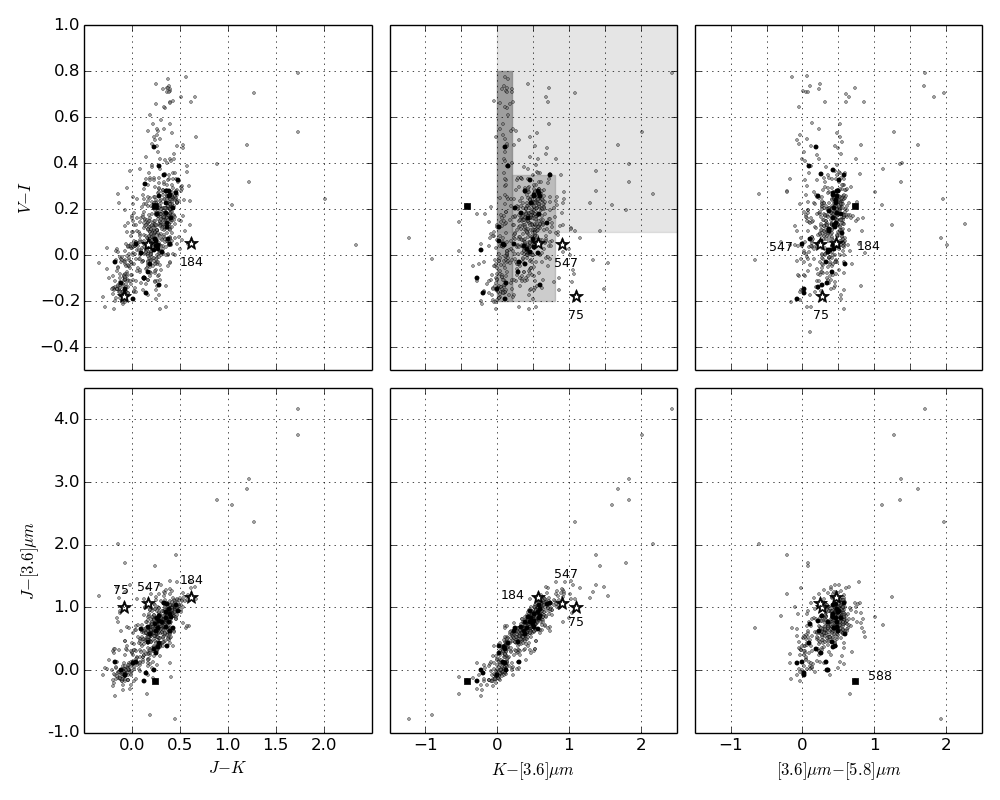}
    \caption{Optical, near- and mid-infrared colour-colour diagrams for the SMC3 sample. BEL confirmed stars are highlighted in darker symbols.
        Large white star symbols are stars with ID = 75, 184, 247. Some stars (including BEL confirmed ones) in the SMC3 sample are missing either IRSF or IRAC photometry. The upper middle panel shows in decreasing shade gray colour, the regions corresponding to the subsets A, B and C, respectively. 
SMC confirmed BEL star ID = 588 (black dark square symbol) is a high mass X-ray binary, which can explain its colours being visibly different from those of CBe stars.}
    \label{fig_cc_smc}
\end{figure*}

Figures \ref{fig_cc_lmc} and \ref{fig_cc_smc} show the optical, near- and mid-infrared colour-colour diagrams
for the LMC3 and SMC3 samples,  highlighting confirmed BEL stars. 
These plots can be compared with similar or related ones in \citet{mennickent02}, \citet{sabogal05}, 
\citet{bonanos11}, \citet{paul12} and \citet{vieira17}. 
Among the Magellanic Be candidates, we distinguish three subsets of data that cluster or locate along some visible sequences simultaneously in different colours.  We broadly define subsets  A, B and C, based on the $V-I$ vs. $K-[3.6] \mu m$ plot, in the upper mid panel of the aforementioned figures, as follows:
\begin{enumerate}
\item[A:] Stars with  $0.0\lesssim K-[3.6] \mu m \lesssim0.2$ and $-0.2\lesssim V-I\lesssim 0.8$.
These stars occupy a vertical sequence locus with 
$[3.6]\mu m-[5.8] \mu m\sim 0$ in the upper right plot and a slightly tilted straight sequence in the upper left one, easily visible for $V-I>0.3$.
\item[B:] Stars with  $0.2\lesssim K-[3.6] \mu m \lesssim0.8$, that occupy an extended locus, having $-0.2\lesssim V-I\lesssim 0.35$ and $J-K<0.5$.
These stars are in the expected location for CBe stars, as can be inferred from \cite{jesush}.
\item[C:] Stars with $K-[3.6]\mu m\gtrsim 0.0$ and $V-I\gtrsim 0.1$, that also have other near- and mid-infrared colours $\gtrsim 0.5-1.0$ and are
not in samples A and B.
\end{enumerate}

The expected locus of CBe stars, our subset B, is very well followed by a large portion of our Magellanic Be candidates as well as our confirmed BEL  stars. We find six LMC confirmed BEL stars in our subset C (ID $= 531, 909, 931, 934, 1368, 1863$), that  are clearly redder than CBe in all of the colours considered, except for one (ID $= 934$) that is exceedingly red in optical and becomes progressively
less red in larger wavelength colours (black dark stars symbols in Figure \ref{fig_cc_lmc}).

Among the SMC confirmed BEL stars, we find only three of them (ID $= 75, 184, 547$) that are redder than or 
in the redder side of CBe ones (large white star symbols in Figure \ref{fig_cc_smc}). 
More specifically, ID=$184$ is redder in $J-K$ and ID=$75, 547$ are redder in $K-[3.6]\mu m$ than CBe stars, and all three of them
are among the reddest BEL stars in $J-[3.6]\mu m$ colours, although these latter values are within the expected for CBe stars.

Table \ref{tab_redbe} lists the individually identified red confirmed BEL stars in the LMC (6) and the SMC (3), that we call {\it red Be stars}, 
indicating our ID (columns 1 and 4), identification and spectral type listed by \citet{reid12} (columns 2 and 3), and by \citet{sheets13} (columns 5 and 6). 
In these references the stars were spectroscopically confirmed as BEL, since the $H_{\alpha}$ line is observed clearly in emission, although the spectra obtained were of low resolution. In this sense, these stars could also be Herbig Ae/Be (HAe/Be), B[e], or other type of emission line B star than CBe. As reported by \citet{review13}, classical Be stars can be confused with HAe/Be stars, which are pre-main-sequence objects still embedded in clouds of gas and dust and 
are accreting material from circumstellar disks that are by products of the star forming processes  \citep{Waters}. CBe stars, on the other hand, are not pre-main-sequence objects. Their circumstellar disks originate from outbursts from the star and not from an external cloud. Typically, infrared and millimetric observations are used to distinguish between CBe and HAe/Be stars.

We performed a spectral energy distribution analysis to infer whether 
the observed infrared excess for these red Be stars is produced by dust, interstellar extinction or
by thermal emission from circumstellar dust (e.g. protoplanetary disk, envelope or both). We found five red Be stars of the LMC with infrared colours expected for HAe/Be. The LMC star $ID = 909$ and the three red  Be stars of the SMC are not located in that region, but in a over-density of blue stars ($B-V<0.1$) in the colour-colour diagram, implying that they could not be HAe/Be.

On the other hand, an additional aspect observed in our results is the difference between the number of red Be stars in the LMC and SMC. It is required to find out if this is indicative of an intrinsic difference between the BEL population of both galaxies,
related for example to metallicity, or if it is caused by some bias/incompleteness in the available data.

In the following, we give a brief description of the four objects called here red Be stars.
The SMC stars ID $= 75, 184$, $547$ (Type-1, Type-2 and Type-4, respectively, in the classification of \citealt{mennickent02}) were reported by \cite{sheets13} in Table 4,  and their spectra are showed in Figure 4 of that work. Those authors could not assure if the objects in their Table 4 reported as OBe (stars with a clear $H_{\alpha}$ line in emission) were CBe or HAe/Be stars, because of their short ages. Although the spectra in their Figure 4 have a low resolution of 570, it is clear the presence of the $H_{\alpha}$ emission line. This is why we called them confirmed BEL stars. However, since in our analysis they cannot be classified as HAe/Be, nor discarded as CBe,  it will be necessary to observe  them in a medium or high resolution spectroscopic multi-epoch follow up to investigate their Be nature. The LMC star ID $= 909$ (Type-3 in the classification of \citealt{mennickent02}) was reported by \citet{reid12} in Table A1. The authors used low resolution spectra for identifying the emission lines and performing the spectral classification. Then, the stars were observed with a medium resolution of 1200 to obtain radial and rotational velocities. Finally, they classified the $H_{\alpha}$ line profiles using  higher resolution spectra of 8600. The spectral classification for star 909 is B3Ve, indicating that it does not present forbidden lines, it is not a supergiant, and shows the $H_{\alpha}$ line in emission. The high resolution of the stellar spectrum allows to say that this is a real emission line and not a spurious detection. Table A1 indicates that this line has a double peaked profile with the absorption characteristic just in the center. This implies that the star is viewed with a mid-inclination angle \citep{reid12}. The rotational velocity obtained in that work for star ID $= 909$ is $328 \pm 16$ km s$^{-1}$, that is in the range of velocities of CBe. On the base of these characteristics, this star could be a Be star of the LMC. However, again, it will be necessary to observe it in a medium or high resolution spectroscopic multi-epoch follow up to confirm its Be nature.  

Having found four red Be stars in the SMC (ID = 75, 184, 547) and LMC (ID = 909), is an important fact that suggests there could be more of these exceptionally red Be stars in these galaxies. Such a high reddening could be an indication of dust, which should 
not exist in the circumstellar environment of CBe stars, due to the high 
effective temperatures of their central stars. However, in case that one or more of these stars were CBe, this result would have important consequences about the physics of circumstellar disks of these stars. Further investigation is required to find the best explanation for the existence of these red Be stars in the Magellanic clouds. In the following section we present some hypotheses to explain  these red colours that exclude the idea of red CBe. These studies require additional observations and analysis.

\section{Alternative explanations for the redder colours of our red B\lowercase{e} stars}\label{sec_al_red_be}

There are many possible explanations for our results than by truly red CBe stars. 
One of them is the possibility of stellar companions, by binarity or blending, causing redder optical and IR colours. High resolution spectra will be needed to confirm the existence of such companion stars. Another possibility is that red Be stars in the MC belong to high mass X-ray binary systems. The SMC confirmed BEL star ID $= 588$ (labeled [M2002] SMC 17703 in \citealt{martayan10}), 
has been identified by \citet{hmxb} as a high mass X-ray binary (No. 45 in their Table A.1). It shows noticeably different colours, some bluer and some redder, than CBe stars  (dark black square in Figure \ref{fig_cc_smc}). However, as far as we know, none of our red Be stars has been reported in a high mass X-ray binary system.

On the other hand, another possible explanation is that a high inclinated disk could produce the red colours showed by the red Be stars. We develop simulations to study this hypothesis. It has been demonstrated \citep{2018MNRAS.476.3555R,2018MNRAS.479.2214G} that a Be disk fed roughly at a constant rate, and for a sufficiently long time (a few to several years, depending on the value of the viscosity parameter $\alpha$), reaches a quasi-steady state in which the density is nearly constant in time.
If the gas temperature is properly taken into consideration, the density profile is typically a complicated function of the distance from the star \citep{2008ApJ...684.1374C}. However, a usual approximation is to consider the temperature of the gas to follow a power-law with the radial distance, in which case the density profile assumes a power-law form given by
\begin{equation}
\rho(R,z) = \frac{\Sigma_0}{\left(2\pi\right)^\frac{1}{2}H_0}\left(\frac{R}{R_\mathrm{eq}}\right)^{-n}e^{-\frac{z^2}{2H^2}}\,,
\end{equation}
where
\begin{equation}
H\left(R\right) = H_0\left(\frac{R}{R_\mathrm{eq}}\right)^\beta\,,~\beta=\frac{3}{2}\,,
\end{equation}
{is the disk scale height, and
\begin{equation}
H_0^2 = \frac{c_s^2}{v_\mathrm{orb}^2}R_\mathrm{eq}^2 = \frac{kT_0}{\mu m_H}\frac{R_\mathrm{eq}^3}{GM}\,.
\end{equation}

For an isothermal disk, we must have $n = 7/2$ \citep{2011IAUS..272..325C}. Typical values of $n$, however, are between $2.5$ and $5$ \citep{2017MNRAS.464.3071V}. In the recent years, this simple steady-state viscous decretion disk has been successful in describing the main observed features of individual Be disks \citep{2006ApJ...652.1617C,2007ApJ...671L..49C,2008ApJ...687..598J,2009A&A...504..915C,2015A&A...584A..85K} and samples of Be stars \citep{2010ApJS..187..228S,2011ApJ...729...17T,2017MNRAS.464.3071V}. We used the radiative transfer code {\tt HDUST} \citep{2004ApJ...604..238C,2006ApJ...639.1081C,2008ApJ...684.1374C} in order to estimate the locii of Be stars in the colour-colour diagrams presented in this work. From a grid of Be star+disk models, we generated a population of models of Be stars with parameters distributed as follows: $n$ is uniformly distributed between $3$ to $4.5$, $\Sigma_0$ is uniformly distributed from $0$ to $4\,\mathrm{g\,cm^{-2}}$, $M$ is distributed from $4.2\,M_\odot$ to $20\,M_\odot$ roughly in accordance with the mass distribution of Be stars shown in  Fig.\,10 from \citet{2018MNRAS.476.3555R}; $W$ is normally distributed, with mean $\left\langle W\right\rangle = 0.81$ and standard deviation $\sigma_W = 0.12$ \citep{2006A&A...459..137R}, and $\cos i$ is uniformly distributed between $0$ and $1$. All stellar models were interpolated from the grids of \citet{2014A&A...566A..21G}, assuming stars at the main sequence, with central hydrogen fraction given by $X_\mathrm{c}=0.3$. The generated population of Be star models should mimic the population of Be star candidates that are found using the variability criterion of \citet{mennickent02} and \citet{sabogal05}.

\begin{figure}
\includegraphics[width=\columnwidth]{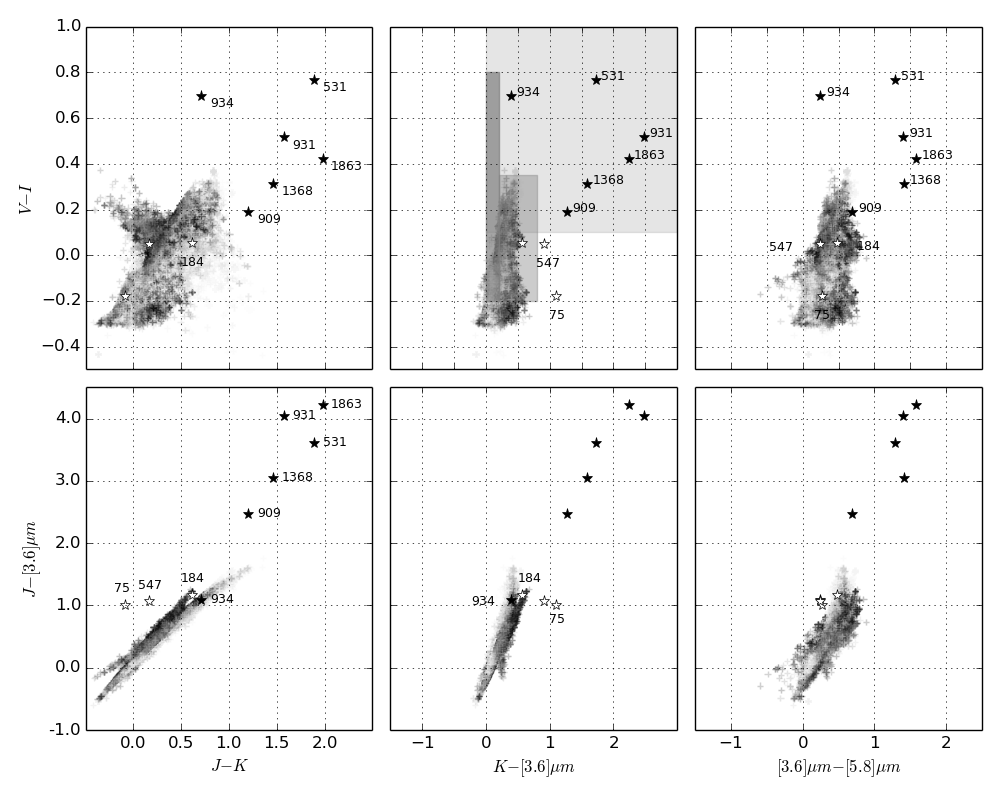}
\caption{Theoretical population of Be stars. The colour of the points reflects the value of disk inclination $\cos(i)$: 
lighter shade is closer to 0 (edge-on) and darker shade in closer to 1 (pole-on).
Pole-on disk stars tend to be redder, yet no alignment configuration can explain the observed red colours of our red Be stars.
Red BEL stars are plotted with star symbols (black for the LMC, white for the SMC).}
\label{Fig10mod}
\end{figure}

In Fig.\,\ref{Fig10mod}, we plot the obtained theoretical population of Be stars, which falls in the region B (the expected locus of Be stars). The colour of the points reflect the values of $\cos i$. Thus, more pole-on stars tend to be redder. No alignment configuration, however, is capable of producing the observed red colours of our red Be stars. Hence, the standard models cannot explain the Be star candidates of regions A and C, and, particularly, the red Be stars. If future observations indeed confirm that any of our red Be stars is, in fact, a CBe star, it could be the fact that its gaseous circumstellar disk is much denser than any other previously found ($\Sigma_0 \gg 4\,\mathrm{g\,cm^{-2}}$). Such a dense disk, with the typical values of $\alpha$ viscosity parameters derived for Be stars \citep{2018MNRAS.476.3555R,2018MNRAS.479.2214G}, 
would mean that viscous heating would not be negligible as compared to irradiated heating \citep{1999A&A...348..512P} 
and the disk would contain abnormally higher temperature regions. Also, this disk would be removing much more angular momentum from the parent star, as compared to the typical values \citep{2018MNRAS.476.3555R,2018MNRAS.479.2214G}, with consequences for the stellar evolution of Be stars.

On the other hand, in a study of Galactic Be stars showing far-infrared excess, \citet{Mirosch2000} found 9 extended sources, linked probably to infrared cirrus emission, i.e, patches of overdense interstellar medium heated by nearby stars \citep{Vanburen88}. On this basis, it is possible that another explanation for our red Be stars can be that they are extended sources embedded in expanding bubbles. High resolution images in far-IR are needed to prove this hypothesis. 

Finally, the IR colours and $H_{\alpha}$ emission lines observed in our red Be stars, can be explained also if they are B[e] stars: BEL stars that show a large IR excess caused by a hot circumstellar dust \citep{Mirosch2007}. In their optical spectra there are forbidden emission lines as [$\mathrm{O}_\mathrm{I}$], [$\mathrm{Fe}_\mathrm{II}$], [$\mathrm{N}_\mathrm{II}$] and [$\mathrm{O}_\mathrm{III}$]; and also permitted lines as the Balmer and $\mathrm{Fe}_\mathrm{II}$ (\citealt{Allen76}, \citealt{Lamers}).  These forbidden emission lines and dust-type infrared excess distinguish  B[e] from CBe stars. 

The B[e] phenomenon has been detected in supergiant (sgB[e]), pre-main sequence (HAe/B[e]), symbiotic (SymB[e]) and FS CMa stars, as also in compact planetary nebulae (cPNB[e]). Some of FS CMa stars could be transitional objects between binary Be and symbiotic stars \citep{Mirosch2007}. The B[e] phenomenon can be studied with different techniques: multi-epoch high-resolution spectroscopy (with high S/N ratio), multicolour photometry and spectropolarimetry. In order to study if the MC red Be stars are certainly B[e], near- and mid-IR imaging, interferometry or middle resolution spectroscopy should be able to confirm if the emission reported by their low resolution spectra is not spurious. Also, it may detect forbidden lines (e.g.  [$\mathrm{O}_\mathrm{I}$] and [$\mathrm{Ca}_\mathrm{II}$]). In particular, for LMC ID $= 909$ star, the single epoch high resolution spectrum for this star reported by \citet{reid12}, allowed to establish its spectral type and the lack of forbiden lines. Multi-epoch optical and infrared spectroscopy studies can confirm these results. 

At last, a chance alignment within the angular resolution of OGLE and GAIA is not impossible but considered improbable
as all four red Be stars have very similar  OGLE $V$ and GAIA $G$ magnitudes, as shown by the whole sample distribution
in Figure \ref{fig_histo_dtheta_dj}.

The studies mentioned above should be carried out in order to establish the nature of our red Be stars. However, in case of not getting a conclusive result that can explain the observed red excess, perhaps a new type of objects, the red Be stars, will be waiting for an explanation of their  astrophysical features and behaviour.

\begin{table}
	\centering
	\caption{Different samples of Be candidates towards the Magellanic Clouds used in this investigation.}
	\label{tab_all.samples}
	\begin{tabular}{crr} \hline
Sample name  & LMC &  SMC \\ \hline
Original Be candidates &  2446 & 1019 \\
Clean Be candidates  &  2393 & 1004 \\
Matched with GAIA DR2 & 2393 & 1004 \\
With $\varpi, \mu_\alpha\cos{\delta}$  and $\mu_\delta$ & 2357 & 994 \\
Q90 &  2099 & 890 \\
LMC1/SMC1  & 1856 & 872 \\
LMC2/SMC2 &  2116 & 976 \\
LMC3/SMC3 & 2109  & 974 \\
BEL confirmed stars & 41 & 56 \\
Matched with IRSF & 1521 & 737 \\
Matched with SAGE IRAC & 1882 & 869 \\
  \hline
	\end{tabular}
\end{table}

 \begin{table}
	\centering
	\caption{Magellanic red Be stars: confirmed BEL  stars with redder colours than those of CBe. 
	Stars with an asterisk could not be classified as HAe/Be.}
	\label{tab_redbe}
	\begin{tabular}{c  c  c  c c  c  c} \hline
 LMC   & IDR   &  ST     & & SMC  & IDS    & ST  \\ \hline
  531  & 92   &  B0IIIe & & *75  & B103  & OBe  \\
  *909  & 1386 &  B3Ve   & & *184 & B081  & OBe \\  
  931  & 1265 &  B8Ve   & & *547 & B049  & OBe \\
  934  & 606  &  B1Ve   & &  \\
  1368 & 684  &  B7Ve   & &  \\
  1863 & 1335 &  B1Ve   & &  \\
  \hline	
    \end{tabular}
\end{table}

\section{Conclusions}
\begin{itemize}
\item A proper motion investigation for a sample of Be candidate stars towards the Magellanic Clouds,
using data from the GAIA DR2 catalogue has been done, which confirmed a previous result from \citet{vieira17},
where a contaminant Galactic foreground population with redder colours in $B-V$ and $V-I$ was found.
\item Yet, thanks to the precision and accuracy of GAIA parallaxes and proper motions, we found that some red
Be candidates do belong to the Magellanic Clouds.
\item Within the 2393 and 1004 Be candidates in the LMC and SMC, respectively, we found 41 and 56 already spectroscopically
confirmed B emission line stars, though those marked as Type-2 or 3 must be examined more carefully. 
All but one of these B emission line confirmed stars belong to the Magellanic Clouds.
That one star was not measured by GAIA DR2, therefore its membership to the MCs could not be assessed.
\item Most of our confirmed B emission line  stars are Type-4, but there is a larger proportion of Type-1 vs Type-4, in the SMC
than in the LMC. This is consistent with an idea put forward by \citet{sabogal05}: eruptive Be stars
are more common in metal-poorer SMC. 
\item One LMC confirmed B emission line star shows significantly redder optical, near- and mid-infrared colours than what has been typically measured for Classical Be stars. For the SMC, three confirmed B emission line  stars show reddish colours but not consistently in all bands. We called these four stars red Be stars, based on the fact that they are Be candidates with the $H_{\alpha}$ line in emission and colours redder than these of CBe. Some of our Be candidates, in both LMC and SMC, follow the overall redder colours trend, suggesting the existence of more Magellanic red Be stars. 
\item The abnormal reddening of red Be stars deserves further investigation in order to clarify which kind of objects they are. Possible explanations for these red Be stars are stellar companions, remnant dust from previous evolution, bubbles in the interstellar medium,  B[e] stars, among others. Further detailed investigation needs to be carried out as a follow up of this study in order to confirm or discard these hypotheses and the Be nature of these stars, in order to find the true explanation for them. If they were CBe stars their red colours would imply that our understanding of the physics of the CBe stars and their circumstellar decretion disks would need to be reviewed, in the sense that until now, these disks have been considered free of dust due to the way in which they are formed. Infrared spectrocopic observations of these stars could be useful to discard the existence of dust in the disks.
\end{itemize}

%


\section*{Acknowledgements}
This investigation has made use of the following software and programming language: TOPCAT \citep{topcat} and 
Python (Python Software Foundation, https://www.python.org/). This work has made use of data from the European Space Agency (ESA) mission
{\it Gaia} (\url{https://www.cosmos.esa.int/gaia}), processed by the {\it Gaia}
Data Processing and Analysis Consortium (DPAC,
\url{https://www.cosmos.esa.int/web/gaia/dpac/consortium}). Funding for the DPAC
has been provided by national institutions, in particular the institutions
participating in the {\it Gaia} Multilateral Agreement. 
This work has made use of the BeSS database, operated at LESIA, Observatoire de Meudon, France: 
\url{http://basebe.obspm.fr}. 
We thank to the anonymous referees for their useful comments and suggestions that have improved this work. 
Vieira K. wants to thank the Department of Physics at Universidad
de los Andes, Bogot\'a, Colombia for supporting two research stays at their facilities, that made possible
this investigation. 
García-Varela A. and Sabogal B. acknowledge financial support to this
work given by Fondo de Investigaciones de la Facultad de Ciencias
de la Universidad de los Andes, Colombia, through Convocatoria 2016-2 
para la Financiación de proyectos de Investigación Categoría: Profesores 
de Planta, proyecto “Classification of variable stars using machine learning techniques"
and Programa de Investigación código INV-2019-84-1857.
J.H. acknowledges support from the National Research Council of M\'exico 
(CONACyT) project No. 86372 and the PAPIIT UNAM project IA102921.

\section*{Data Availability Statement}

The data underlying this article are available in its online supplementary material.




\begin{thebibliography}{99}
\bibitem[\protect\citeauthoryear{Allen \& Swings}{1976}]{Allen76} Allen, D. A., Swings, H. P. 1976, \aap, 47, 293
\bibitem[\protect\citeauthoryear{Allen}{2000}]{allen} Allen C.W. 2000, in Cox A., ed., Allen's Astrophysical Quantities, 4th edition Springer-Verlag, New York
\bibitem[\protect\citeauthoryear{Arenou et al.}{2018}]{arenou18} Arenou F. et al., 2018, 616, A17
\bibitem[\protect\citeauthoryear{Bonanos et al.}{2011}]{bonanos11} Bonanos A. Z. et al., 2011, IAUS 272, 264
\bibitem[\protect\citeauthoryear{Carciofi et~al.}{2004}]{2004ApJ...604..238C} Carciofi, A.~C., Bjorkman, J.~E., \& Magalh{\~a}es, A.~M. 2004, \apj, 604, 238
\bibitem[\protect\citeauthoryear{Carciofi et~al.}{2006}]{2006ApJ...652.1617C} Carciofi, A.~C., Miroshnichenko, A.~S., Kusakin, A.~V., et al. 2006,  \apj, 652, 1617
\bibitem[\protect\citeauthoryear{Carciofi \& Bjorkman}{2006}]{2006ApJ...639.1081C} Carciofi, A.~C., \& Bjorkman, J.~E., 2006, \apj, 639, 1081
\bibitem[\protect\citeauthoryear{Carciofi et~al.}{2007}]{2007ApJ...671L..49C} Carciofi, A.~C., Magalh{\~a}es, A.~M., Leister, N.~V., Bjorkman, J.~E.,
  \& Levenhagen, R.~S., 2007, \apjl, 671, L49
\bibitem[\protect\citeauthoryear{Carciofi \& Bjorkman}{2008}]{2008ApJ...684.1374C} Carciofi, A.~C., \& Bjorkman, J.~E.,2008, \apj, 684, 1374
\bibitem[\protect\citeauthoryear{Carciofi et~al.}{2009}]{2009A&A...504..915C} Carciofi, A.~C., Okazaki, A.~T., Le Bouquin, J. B., et~al., 2009, \aap, 504, 915
\bibitem[\protect\citeauthoryear{Carciofi}{2011}]{2011IAUS..272..325C}
Carciofi, A.~C. 2011, in IAU Symposium, Vol. 272, Active OB Stars: Structure,  Evolution, Mass Loss, and Critical Limits, ed. C.~{Neiner}, G.~{Wade},  G.~{Meynet}, \& G.~{Peters}, 325
\bibitem[\protect\citeauthoryear{Cieslinski et al.}{2013}]{cieslinski13} Cieslinksi D., Diaz M., Mennickent R. E., Kolaczkowski Z, Pereira C., 2013, IBVS 6088 (C13)
\bibitem[\protect\citeauthoryear{Collins}{1987}]{collins} Collins, G. W. II. (1987) In IAU Colloq. 92: Physics of Be stars. p.3
\bibitem[\protect\citeauthoryear{Feigelson \& Babu}{2012}]{Fei} Feigelson E. D., Babu G. J., 2012, ``Modern Statistical Methods for Astronomy'', Cambrige University Press, New York
\bibitem[\protect\citeauthoryear{GAIA Collaboration}{2016}]{gaia} GAIA Collaboration, 2016, \aap, 595, A1
\bibitem[\protect\citeauthoryear{GAIA Collaboration}{2018a}]{gaiadr2} GAIA Collaboration, 2018a, \aap, 616, A1 
\bibitem[\protect\citeauthoryear{GAIA Collaboration}{2018b}]{gaiadr2MCpm} GAIA Collaboration, 2018b, \aap, 616, A12 
\bibitem[\protect\citeauthoryear{Galad\'i-Enr\'iquez et al.}{1998}]{galadi} Galad\'i-Enr\'iquez D., Jordi C., Trullols E., 1998, \aap, 337, 125
\bibitem[\protect\citeauthoryear{Georgy et al.}{2014}]{2014A&A...566A..21G}
Georgy, C., Granada, A., Ekstr{\"o}m, S., et al., 2014, \aap, 566, A21
\bibitem[\protect\citeauthoryear{Ghoreyshi et al.}{2018}]{2018MNRAS.479.2214G} Ghoreyshi, M.~R., Carciofi, A.~C., R{\'\i}mulo, L.~R., et~al. 2018,  \mnras, 479, 2214
\bibitem[\protect\citeauthoryear{Girard et al.}{2010}]{spm4} Girard T.~M. et al., 2011, \aj, 142, 15
\bibitem[\protect\citeauthoryear{Gordon et al.}{2011}]{gordon11} Gordon K. D. et al., 2011, \aj, 142, 102
\bibitem[\protect\citeauthoryear{Gyuk et al.}{2000}]{gyuk00} Gyuk G., Dalal N., Griest K., 2000, \aj, 535, 90
\bibitem[\protect\citeauthoryear{Haberl \& Sturm}{2016}]{hmxb} Haberl F., Sturm R., 2016, \aap, 586, A81
\bibitem[\protect\citeauthoryear{Heiberger \& Holland}{2015}]{Hei} Heiberger R. M., Holland B., 2012, ``Statistical Analysis and Data Display'', second ed., Springer, New York 
\bibitem[\protect\citeauthoryear{Hern\'andez et al.}{2005}]{jesush} Hern\'andez J., Calvet N., Hartmann L., Brice\~no C., Sicilia-Aguilar A., Berlind P., 2005, \aj, 129, 856
\bibitem[\protect\citeauthoryear{Heumann et al.}{2016}]{Heu} Heumann C., Schomaker G. J., Shalabh 2016, ``Introduction to Statistics and Data Analysis'', Springer, Switzerland
\bibitem[\protect\citeauthoryear{Jones et~al.}{2008}]{2008ApJ...687..598J}
Jones, C.~E., Tycner, C., Sigut, T.~A.~A., Benson, J.~A., \& Hutter,
  D.~J., 2008, \apj, 687, 598
\bibitem[\protect\citeauthoryear{Kato et al.}{2007}]{kato07} Kato D. et al., 2007, \pasj, 59, 615
\bibitem[\protect\citeauthoryear{Klement et~al.}{2015}]{2015A&A...584A..85K}
Klement, R., Carciofi, A.~C., Rivinius, T., et~al., 2015, \aap, 584, A85
\bibitem[\protect\citeauthoryear{Lamers et al.}{1998}]{Lamers} Lamers, H. J.~G.~L.~M., Zickgraf, F.-J., de Winter, D., Houziaux, L.,
  \& Zorec, J., 1998, \aap, 340, 117
\bibitem[\protect\citeauthoryear{Luri et al.}{2018}]{luri} Luri X. et al., 2018, \aap, 616, A9
\bibitem[\protect\citeauthoryear{Luri et al.}{2019}]{luri19} Luri X. on behalf of GAIA DPAC, 2019, 
Highlights on Spanish Astrophysics X, Proceedings of the XIII Scientific Meeting of the Spanish
Astronomical Society held on July 16-20, 2018, in Salamanca, Spain. B. Montesinos, A. Asensio Ramos,
F. Buitrago, R. Schödel, E. Villaver, S. Pérez-Hoyos, I. Ordóñez-Etxeberria (eds.), 16
\bibitem[\protect\citeauthoryear{Martayan et al.}{2007}]{martayan07} Martayan C., Fremat Y., Hubert A.-M., Floquet M., Zorec J., Neiner C.,2007, \aap, 462, 683
\bibitem[\protect\citeauthoryear{Martayan et al.}{2010}]{martayan10} Martayan C., Baade D., Fabregat J., 2010, \aap, 509, A11
\bibitem[\protect\citeauthoryear{Meixner et al.}{2006}]{meixner06} Meixner M. et al., 2006, \aj, 132, 2268
\bibitem[\protect\citeauthoryear{Mennickent et al.}{2002}]{mennickent02} Mennickent R.~E., Pietrzy\'nski G., Gieren W., Szewczyk O., 2002, \aap, 393, 887
\bibitem[\protect\citeauthoryear{Mennickent et al.}{2003}]{mennickent03} Mennickent R.~E., Pietrzy\'nski G., Diaz, M., Gieren W., 2003, \aap, 399, L47
\bibitem[\protect\citeauthoryear{Mennickent et al.}{2009}]{mennickent09} Mennickent R.~E., Sabogal B., Granada A., Cidale L., 2009, \pasp, 121, 125
\bibitem[\protect\citeauthoryear{Miroshnichenko \& Bjorkman}{2000}]{Mirosch2000} Miroshnichenko, A. S., Bjorkman, K. S., 2000, IAU Colloq. 175: The Be Phenomenon in Early-Type Stars (ASP Conf. Ser. 214), ed. M. A. Smith, H. F. Henrichs, \& J. Fabregat (San Francisco, CA: ASP), 484
\bibitem[\protect\citeauthoryear{Miroshnichenko}{2007}]{Mirosch2007} Miroshnichenko, A. S. 2007, \apj, 667, 497
\bibitem[\protect\citeauthoryear{Patel \& Read}{1996}]{booknorm} Patel J. K., Read C. B., 1996, ``Handbook of the Normal Distribution'', Marcel Dekker Inc., New York 
\bibitem[\protect\citeauthoryear{Porter}{1999}]{1999A&A...348..512P}
Porter, J.~M. 1999, \aap, 348, 512
\bibitem[\protect\citeauthoryear{Paul et al.}{2012}]{paul12} Paul K.~T., Subramaniam A., Mathew B., Mennickent R. E., Sabogal B., 2012, \mnras, 421, 3622 
\bibitem[\protect\citeauthoryear{Reid et al.}{2012}]{reid12} Reid W. A., Parker Q. A., 2012, \mnras, 425, 355
\bibitem[\protect\citeauthoryear{R{\'\i}mulo et~al.}{2018}]{2018MNRAS.476.3555R}
R{\'\i}mulo, L.~R., Carciofi, A.~C., Vieira, R.~G., et~al., 2018,
  \mnras, 476, 3555
\bibitem[\protect\citeauthoryear{Rivinius et~al.}{2006}]{2006A&A...459..137R} Rivinius, T., {\v S}tefl, S., \& Baade, D. 2006, \aap, 459, 137
\bibitem[\protect\citeauthoryear{Rivinius et al.}{2013}]{review13} Rivinius T., Carciofi A. C., Martayan, C., 2013, \aapr, 21, 69
\bibitem[\protect\citeauthoryear{Sabogal et al.}{2005}]{sabogal05} Sabogal B.~E., Mennickent R. E., Pietrzinsky G., Gieren W., 2005, \mnras, 361, 1055
\bibitem[\protect\citeauthoryear{Sheets et al.}{2013}]{sheets13} Sheets H. et al., 2013, \aap, 711, 111
\bibitem[\protect\citeauthoryear{Stassun \& Torres}{2018}]{stassun18} Stassun K.G., Torres G., 2018, \apj, 862, 61
\bibitem[\protect\citeauthoryear{Taylor M.~B.}{2005}]{topcat} Taylor M.~B., 2005, \aspc, 347, 29
\bibitem[\protect\citeauthoryear{Silaj et~al.}{2010}]{2010ApJS..187..228S}
Silaj, J., Jones, C.~E., Tycner, C., Sigut, T.~A.~A., \& Smith, A.~D.,
  2010, \apjs, 187, 228
\bibitem[\protect\citeauthoryear{Touhami et~al.}{2011}]{2011ApJ...729...17T}
Touhami, Y., Gies, D.~R., \& Schaefer, G.~H., 2011, \apj, 729, 17
\bibitem[\protect\citeauthoryear{van Buren \& McCray}{1988}]{Vanburen88} van Buren, D. \& McCray, R., 1988, \apjl, 329, L93
\bibitem[\protect\citeauthoryear{Vasilevskis et al.}{1958}]{vasilevskis} Vasilevskis S., Klemola A., Preston G., 1958, \aj, 63,387
\bibitem[\protect\citeauthoryear{Vieira  et al.}{2010}]{vieira10} Vieira K. et al., 2010, \aj, 140, 1934
\bibitem[\protect\citeauthoryear{Vieira  et al.}{2017}]{vieira17} Vieira K., Garc\'ia-Varela A., Sabogal B., 2017, \mnras, 469, 4175
\bibitem[\protect\citeauthoryear{Vieira R.~G. et~al.}{2017}]{2017MNRAS.464.3071V}
Vieira, R.~G., Carciofi, A.~C., Bjorkman, J.~E., et~al., 2017, \mnras,
  464, 3071
\bibitem[\protect\citeauthoryear{Waters \& Waelkens}{1998}]{Waters} Waters, L.~B.~F.~M., \& Waelkens, C., 1998, \araa, 36, 233
\bibitem[\protect\citeauthoryear{Weisstein E. W.}{}]{statsbook} Weisstein E. W.,``Chi Distribution'', from MathWorld -- A Wolfram Web Resource \url{http://mathworld.wolfram.com/ChiDistribution.html}
\bibitem[\protect\citeauthoryear{Zhang  et al.}{2005}]{zhang05} Zhang P., Chen P.~S., \& Yang H.~T., 2005, NewA, 10, 325 
\bibitem[\protect\citeauthoryear{Zinn  et al.}{2019}]{zinn19} Zinn J.C., Pinsonneault M. H., Huber D., Stello D., 2019, \aj, 878, 136
\end{thebibliography}




\appendix

\section{Tables}
A sample of the LMC data used in this investigation is shown below (Table A1). 
Table A2 shows the same data for our four red Be stars.
Full tables (one for the LMC, one for the SMC and one for the four red Be stars that has stars from both Clouds)
are available as online supplementary material. 

\bsp	

\begin{landscape}
\begin{table}
	\centering
	\caption{Sample of the LMC data used in this investigation. It includes OGLE-II ID and BVI photometry, Variability Type (Type=1.5 is for Type 1/2, i.e. sharing both values);
	GAIA DR2 source ID, coordinates, parallax and proper motions with their errors, and BGR photometry; our posmag\_ranking and $\chi$-value for membership to the Clouds; 
	other names for the BEL confirmed stars only (set to --- when no name was given) as listed in the references in Table \ref{tab_crossbe}; 
	IRSF JHKs photometry; and IRAC [3.6,4.5,5.8,8.0] $\mu m$ photometry for each star.}
	\label{tab_final_cat2}
	\begin{tabular}{ccccccccccccc} 
	This paper &  OGLE-II &  V &   B-V  &  V-I   & Variab.  & GAIA DR2   & $\alpha$ & $\delta$  & $\varpi$ & $\epsilon_{\varpi}$ & $\mu_\alpha\cos(\delta)$ & $\epsilon_{\mu_\alpha\cos(\delta)}$   \\  
	 ID &  ID &  (mag) & (mag)  & (mag)  & Type &  source ID & (deg) & (degrees)  & (mas) & (mas) & (mas yr$^{-1}$) & (mas yr$^{-1}$)   \\  \hline
  1  & 05071018-6910538 & 16.899 & -0.074 & 0.006  & 1.0 & 4661232628032602752 & 76.79218055039 & -69.18158550247 & -0.2078 & 0.0752 & 1.634 & 0.171  \\
  2  & 05123453-6914375 & 14.3   & -0.125 & -0.082 & 1.0 & 4658240753888659456 & 78.14363046873 & -69.24371968086 & -0.0796 & 0.1736 & 2.479 & 0.242  \\
  3  & 05173303-6920189 & 17.478 & 0.048  & 0.129  & 1.0 & 4658192233605799808 & 79.38731184581 & -69.33854593851 & -0.0421 & 0.1078 & 1.848 & 0.17   \\
  4  & 05023318-6921184 & 16.015 & -0.068 & -0.043 & 1.0 & 4655233138520014592 & 75.63790883973 & -69.35504535077 & 0.046   & 0.0403 & 1.894 & 0.077  \\
  5  & 05200092-6941138 & 17.986 & 0.077  & 0.18   & 1.0 & 4658166223321197312 & 80.00338154581 & -69.68713050128 & 0.0552  & 0.1437 & 1.919 & 0.206  \\
  6  & 05200237-6927100 & 16.227 & -0.124 & -0.082 & 1.0 & 4658175435947436288 & 80.00947440419 & -69.45277712703 & -0.0887 & 0.0663 & 2.085 & 0.109 \\
  7  & 05200385-6948340 & 17.07  & -0.23  & 0.007  & 1.0 & 4658140904452708736 & 80.01579285108 & -69.80935793508 &         &        &       &              \\
  8  & 05200424-6925374 & 15.661 & -0.162 & -0.14  & 1.0 & 4658176947831707008 & 80.01727024206 & -69.42705739503 & 0.1395  & 0.0678 & 2.656 & 0.109  \\
  9  & 05200714-6939426 & 14.656 & -0.015 & 0.172  & 1.0 & 4658166399388690688 & 80.02930474084 & -69.66182328727 & -0.046  & 0.0873 & 1.643 & 0.134 \\
  10 & 05223928-6952325 & 16.317 & 0.307  & 0.662  & 1.0 & 4657972399949681920 & 80.66322850573 & -69.87567345694 & 8.0E-4  & 0.0375 & 2.072 & 0.066  \\
     & $\vdots$         &        &        &        &     &                     &                &                 &         &        &       &             \\
  25 & 05173037-6920349 & 16.625 & 0.002  & -0.082 & 2.0 & 4658192203563908992 & 79.37610453534 & -69.34299613095 & -0.1427 & 0.0958 & 1.801 & 0.149  \\
  26 & 05173364-6921328 & 16.356 & -0.04  & -0.065 & 2.0 & 4658191344570530816 & 79.38986687638 & -69.35906390055 & 0.0163  & 0.0596 & 2.28  & 0.085  \\
  27 & 05300135-6928099 & 15.776 & -0.207 & 0.238  & 3.0 & 4658043249748073472 & 82.50530387206 & -69.46940253938 & -0.3151 & 0.2066 & 0.887 & 0.336  \\
  28 & 05390419-7005011 & 16.499 & -0.172 & 0.049  & 3.0 & 4657251021605219968 & 84.76693210448 & -70.08363577666 & -0.0461 & 0.0542 & 2.088 & 0.095  \\
  29 & 05045870-6910069 & 16.546 & 0.043  & 0.292  & 4.0 & 4661244374880374400 & 76.24436119511 & -69.16850244506 & -0.0885 & 0.066  & 2.005 & 0.109  \\
  30 & 05050594-6910449 & 15.431 & -0.02  & 0.215  & 4.0 & 4661244306027427584 & 76.27448372936 & -69.17911023874 & 0.0985  & 0.0469 & 1.788 & 0.087  \\ \hline
\end{tabular}

	\centering
	\contcaption{}
	\label{tab_final_cat2}
	\begin{tabular}{ccccccccccccccc} 
	$\mu_\delta$ & $\epsilon_{\mu_\delta}$ & B & G & R & $posmag\_ranking$ & $\chi$ & BEL confirmed & J &  H &   Ks  &  [3.6] $\mu$m   & [4.5] $\mu$m  & [5.8] $\mu$m   & [8.0] $\mu$m  \\  
	(mas yr$^{-1}$) & (mas yr$^{-1}$) & (mag) & (mag) & (mag) & value & value & other names & (mag) & (mag) &  (mag)  & (mag)   & (mag)  & (mag)   & (mag)  \\  \hline
 -0.006 & 0.182 & 16.763409 & 16.642477 & 16.630665 & 0.160 & 1.857 &        &       &       &       & 17.05    &          &          &         \\
 2.137  & 0.318 & 14.250258 & 14.057608 & 14.143691 & 0.101 & 2.322 &        & 14.6  & 14.67 & 14.71 &          &          &          &         \\
 0.424  & 0.218 & 17.37947  & 17.182137 & 16.957771 & 0.146 & 0.268 &        &       &       &       & 16.454   & 16.05    &          &         \\
 -0.043 & 0.097 & 16.013058 & 15.887163 & 16.052021 & 0.130 & 1.650 &        & 15.99 & 15.95 & 15.94 & 15.674   & 15.475   &          &         \\
 0.013  & 0.293 & 18.011503 & 17.789316 & 17.532957 & 0.154 & 0.493 &        &       &       &       &          &          &          &         \\
 0.251  & 0.169  & 16.270792 & 16.122494 & 16.257172 & 0.141 & 1.021 &        &       &       &       &          & 16.126   &          &         \\
             &            & 17.001675 &           &           & 0.126 &       &        &       &       &       & 14.076   & 14.156   & 13.862   &         \\
 0.862  & 0.158  & 15.637781 & 15.453746 & 15.601172 & 0.139 & 3.221 &        & 15.61 & 15.53 & 15.42 & 15.131   & 14.974   & 14.662   &         \\
 0.221  & 0.195 & 14.799377 & 14.578164 & 14.678681 & 0.204 & 0.642 &        &       &       &       & 13.46    & 13.221   & 13.048   &         \\
 0.304  & 0.081 & 16.198282 & 16.35375  & 15.715758 & 0.188 & 1.129 &        & 15.09 & 14.64 & 14.52 & 14.389   & 14.454   & 14.621   &         \\
          & $\vdots$  &           &       &       &        &       &       &       &          &          &          &         \\
 0.243  & 0.197 & 16.738895 & 16.383902 & 16.215532 & 0.188 & 0.895 &        &       &       &       &          &          &          &         \\
 -0.091 & 0.116 & 16.397356 & 16.263079 & 16.30308  & 0.119 & 2.140 &        & 16.17 & 16.21 & 16.17 & 15.696   & 15.626   &          &         \\
 0.222  & 0.481 & 15.477867 & 15.366205 & 15.004803 & 0.316 & 1.395 & RP$\_786$ &       &       &       & 13.731   & 13.576   & 13.307   & 13.1    \\
 0.566  & 0.108 & 16.632305 & 16.523197 & 16.450397 & 0.217 & 1.279 &        & 15.98 & 15.77 & 15.45 & 14.918   & 14.578   & 14.131   &         \\
 0.013  & 0.146 & 16.55129  & 16.427717 & 16.406654 & 0.105 & 1.126 &        & 16.77 & 16.72 & 16.55 &          &          &          &         \\
 0.126  & 0.111 & 15.432999 & 15.337342 & 15.413625 & 0.095 & 1.721 &        & 15.32 & 15.26 & 14.96 & 14.355   & 14.184   & 13.855   & 13.717  \\ \hline
	\end{tabular}
\end{table}

\end{landscape}
\begin{landscape}


\begin{table}
	\centering
	\caption{Same table as before, but only the four red Be stars are included. ID=909 is in the LMC and ID=75, 184, 547 are in the SMC.}
	\label{tab_final_cat3}
	\begin{tabular}{ccccccccccccc} 
	This paper &  OGLE-II &  V &   B-V  &  V-I   & Variab.  & GAIA DR2   & $\alpha$ & $\delta$  & $\varpi$ & $\epsilon_{\varpi}$ & $\mu_\alpha\cos(\delta)$ & $\epsilon_{\mu_\alpha\cos(\delta)}$   \\  
	 ID &  ID &  (mag) & (mag)  & (mag)  & Type &  source ID & (deg) & (degrees)  & (mas) & (mas) & (mas yr$^{-1}$) & (mas yr$^{-1}$)   \\  \hline
 909 &  05163930-6920482    & 17.21 &  -0.079 & 0.191  &  3 &   4658189351701024896  & 79.16335801642  & -69.34666821498 &  0.2341   &  0.0816   &    1.637  &  0.142   \\ \hline
  75 &  005504.55-724637.3  & 14.156 & -0.111   &  -0.178  & 1   &    4685982806630491648  & 13.76896086484  & -72.7770997258  &  0.0447   &  0.0325   &   0.792  &  0.067  \\   
 184 &  005206.05-725208.7  & 15.961 & -0.148  &  0.054  &  2   &    4685959403340845312  & 13.02528552346  & -72.8692141688  &  -0.055   &  0.0394   &   0.581   & 0.079  \\ 
 547 &  004942.75-731717.7  & 14.453 &  -0.115  & 0.049 &   4    &   4685931679327948800  & 12.42810225874 &  -73.28838419448 &  -0.1164   & 0.0314   &  0.356  &  0.054  \\ \hline
        \end{tabular}

	\centering
	\contcaption{}
	\label{tab_final_cat4}
	\begin{tabular}{ccccccccccccccc} 
	$\mu_\delta$ & $\epsilon_{\mu_\delta}$ & B & G & R & $posmag\_ranking$ & $\chi$ & BEL confirmed & J &  H &   Ks  &  [3.6] $\mu$m   & [4.5] $\mu$m  & [5.8] $\mu$m   & [8.0] $\mu$m  \\  
	(mas yr$^{-1}$) & (mas yr$^{-1}$) & (mag) & (mag) & (mag) & value & value & other names & (mag) & (mag) &  (mag)  & (mag)   & (mag)  & (mag)   & (mag)  \\  \hline
-0.104 &  0.183 &  17.15636   &   16.814718  &  16.611889  & 0.14827046694154847  & 2.2372665627995842 &  SAB899             & 16.49   & 15.95 &   15.29  & 14.026  & 13.592  & 13.335 &  13.088 \\ \hline
-1.274 &  0.052  & 14.16928    &  40.3053      &  14.307247  & 0.12046620512225799  & 1.697884385478025   & smc1-17, B103   &  14.56  &  14.6   &  14.65  & 13.548 &  13.413  & 13.281  & 13.046 \\
-1.111 &  0.064  &  15.766048 &  61.8202       &  15.724222  & 0.30135290880066146  & 1.2536129357415433 &  B081                 & 15.67  &  15.37  &  15.06  & 14.498 &  14.204  & 14.026  & 13.599 \\
-1.096 &  0.051 &  14.462686 &  162.339        & 14.502104  & 0.23370929985365374  & 3.831751691002231   &  B049                 &14.47  &  14.39   & 14.31 &  13.405  & 13.225 &  13.162  & 13.034 \\
\hline
	\end{tabular}
\end{table}

\end{landscape}


\label{lastpage}
\end{document}